\documentstyle[12pt,epsf,epsfig]{article}
\def\beq{\begin{equation}}   \def\eeq{\end{equation}}
\def\lsim{\mathrel{\rlap{\lower3pt\hbox{\hskip0pt$\sim$}}
    \raise1pt\hbox{$<$}}}         %less than or approx. symbol
\def\gsim{\mathrel{\rlap{\lower4pt\hbox{\hskip1pt$\sim$}}
    \raise1pt\hbox{$>$}}}         %greater than or approx. symbol
\def\simlt{\mathrel{\raise.3ex\hbox{$<$\kern-.75em\lower1ex\hbox{$\sim$}}}}
\def\simgt{\mathrel{\raise.3ex\hbox{$>$\kern-.75em\lower1ex\hbox{$\sim$}}}}
\begin{document}
\begin{titlepage}
\begin{flushright}
hep-ph/0306240\\
FTPI--MINN--03/14\\
UMN--TH--2203/04\\
CERN--TH/2003--121\\
\end{flushright}
\begin{center}
\baselineskip25pt

\vspace{1cm}

{\Large\bf CP Violation in Supersymmetric $U(1)'$ Models}

\vspace{1cm}

{\sc Durmu{\c s} A. Demir$^{1}$} and {\sc Lisa L. Everett$^{2}$}
\vspace{0.3cm}  

$^{1}$ William I. Fine Theoretical Physics Institute,\\
University of Minnesota, Minneapolis, MN 55455, USA\\
\vspace{0.2cm}
$^{2}$ CERN, TH Division, CH-1211 Geneva 23, Switzerland
\end{center}
\vspace{1cm}
\begin{abstract}

The supersymmetric CP problem is studied within superstring-motivated
extensions of the MSSM with an additional $U(1)'$ gauge symmetry broken at
the TeV scale.  This class of models offers an attractive solution to the
$\mu$ problem of the MSSM, in which $U(1)'$ gauge invariance forbids the
bare $\mu$ term, but an effective $\mu$ parameter is generated by the
vacuum expectation value of a Standard Model singlet $S$ which has
superpotential coupling of the form $S H_u H_d$ to the electroweak Higgs
doublets. The effective $\mu$ parameter is thus dynamically determined as
a function of the soft supersymmetry breaking parameters, and can be
complex if the soft parameters have nontrivial CP-violating phases.  We
examine the phenomenological constraints on the reparameterization
invariant phase combinations within this framework, and find that the
supersymmetric CP problem can be greatly alleviated in models in which the
phase of the $SU(2)$ gaugino mass parameter is aligned with the soft
trilinear scalar mass parameter associated with the $S H_u H_d$ coupling.
We also study how the phases filter into the Higgs sector, and find that
while the Higgs sector conserves CP at the renormalizable level to all
orders of perturbation theory, CP violation can enter at the
nonrenormalizable level at one-loop order.  In the majority of the
parameter space, the lightest Higgs boson remains essentially CP even
but the heavier Higgs bosons can exhibit large CP-violating mixings, 
similar to the CP-violating MSSM with large $\mu$ parameter.

\end{abstract}

\end{titlepage}
\section{Introduction}

While phenomenological models with low energy supersymmetry (SUSY) are
arguably the best candidates for physics beyond the Standard Model (SM),
they typically include a large number of parameters associated with the
soft supersymmetry breaking sector.  For example, the minimal
supersymmetric standard model (MSSM), which has two Higgs doublets and 
conserved R-parity, contains 105 new parameters
\cite{Dimopoulos:1995ju}, including the bilinear Higgs
superpotential parameter $\mu$ and the soft SUSY breaking
parameters (this counting does not include the gravitino mass and
coupling). The parameter count generically increases if such
SUSY models are extended beyond the minimal gauge structure and
particle content of the MSSM (unless symmetry relations exist in the
theory which relate subsets of parameters). 
Many of these new parameters are phases, which both provide new sources of
CP violation and modify the amplitudes for CP-conserving processes.  Even
if certain sectors of the theory exhibit no CP violation at
tree level (e.g. if the relevant phases can be
eliminated by global phase rotations), the phases can leak into such
sectors of the theory at the loop level and have an impact on collider 
phenomenology and cosmology.

In contrast to the SM, in which the only source of CP
violation is present in the CKM matrix and thus is intimately tied to
flavor physics, CP-violating phases within SUSY models can occur in both
flavor-conserving and flavor-changing couplings.  The phases of the 
flavor-conserving couplings (which have no analogue in the SM) are of
particular interest because they can have significant phenomenological
implications which can be studied without knowledge of the origin of
intergenerational mixing.  In the MSSM, these phases are given by  
reparameterization invariant combinations of the phases of
the gaugino mass parameters $M_a$ ($a=1,2,3$), the trilinear couplings
$A_{f}$, the $\mu$ parameter, and the Higgs bilinear coupling $b\equiv \mu
B$. However, not all of these phases are physical; after utilizing the
$U(1)_{PQ}$ and $U(1)_R$ global symmetries of the MSSM, the 
reparameterization invariant phase combinations are 
$\theta_{f}=\phi_{\mu}+\phi_{A_f}-\phi_{b}$ and
$\theta_{a}=\phi_{\mu}+\phi_{M_a}-\phi_{b}$ (in self-evident notation).

Such phases have traditionally been assumed to be small due to what is
known as the supersymmetric CP problem: the experimental upper limits on
the electric dipole moments (EDMs) of the electron, neutron, and certain
atoms individually constrain the phases to be less than $O(10^{-2})$ (for 
sparticle masses consistent with naturalness) 
\cite{oldedm,Dugan:1984qf,newedm}.  
However, recent studies have shown that EDM bounds can be satisfied
without requiring all reparameterization invariant phase combinations to
be small, if either (i)  certain cancellations exist between different EDM
contributions 
\cite{ibrahimnath,Brhlik:1998zn,Pokorski:1999hz,Barger:2001nu,Abel:2001vy}, 
or (iii) the
sparticles of the first and second families have multi-TeV masses
\cite{heavy12}.\footnote{The EDM bounds are more 
difficult to satisfy in both of these scenarios when $\tan\beta$ (the 
ratio of electroweak Higgs VEVs) is large. Not only are cancellations 
in the one-loop EDMs more difficult to achieve, but certain two-loop 
contributions are then enhanced \cite{twoloop,largetanbetaEDM} which do 
not decouple when the sfermions are heavy.  In part for this reason, we 
will restrict our attention in this paper to the small $\tan\beta$ 
regime.}  Even within each of these scenarios, the particularly strong 
constraints arising from the atomic EDMs \cite{Falk:1999tm} lead to a 
general upper bound of $\simlt O(10^{-3})$ on the reparameterization 
invariant phase present in the chargino sector ($\theta_2$ in our 
notation), while the other phases are comparatively unconstrained 
\cite{Barger:2001nu}. These constraints will be discussed in detail later 
in the 
paper; for now, it is worth noting that this ``CP hierarchy problem" is an 
intriguing issue to be addressed within models of the soft parameters 
which include CP violation.\footnote{However, there are unavoidable 
theoretical uncertainties involved in the determination of the hadronic 
EDMs and the atomic EDMs (see e.g. \cite{Pilaftsis:2002fe,Demir:2002gg} 
for discussions). These uncertainties are particularly problematic for the 
mercury EDM, which yields the strongest constraints on the SUSY phases. 
For this reason, there are disagreements in the literature over how to 
include this bound. Here we take a conservative approach by 
including the Hg EDM constraint.}

Of course, if the reparameterization invariant phases are 
sizeable, they can have important phenomenological consequences. Within 
the MSSM, one of the examples in which these phases can have a significant 
impact is the Higgs sector.  As is well known, the MSSM Higgs sector 
conserves CP at tree level.  However,
radiative corrections involving the SM fields and their superpartners
(with the dominant effects typically due to top and stop loops)  have a
substantial impact on Higgs masses and mixings. For example, the 
one-loop radiative corrections substantially elevate the tree-level 
theoretical upper bound of $M_Z$ on the mass of the lightest Higgs boson 
\cite{radiative1}; these results have been 
improved by utilizing complete one-loop on-shell 
renormalization\cite{on-shell}, renormalization
group methods \cite{renorm}, diagrammatic methods with leading order QCD
corrections \cite{diag}, two-loop on-shell renormalization
\cite{Heinemeyer:1998np}, and complete two-loop effective potential
\cite{effpot}. Indeed, if the radiative corrections 
include a nontrivial dependence on phases, the Higgs potential violates CP 
explicitly at one-loop.  The Higgs mass eigenstates then no longer have 
definite CP properties, which leads to important implications for Higgs 
production and decay \cite{pilaftsis,refined,CPX,CPXX}.

The MSSM offers a minimal framework for stabilizing the Higgs sector
against quadratic divergences. However, it is well known that the MSSM has
a hierarchy problem with respect to the scale of the superpotential $\mu$
parameter \cite{muprob}, which has a natural scale of $O(M_{GUT})$, and
the electroweak scale.  An elegant framework in which to address this
``$\mu$ problem" is to generate the $\mu$ parameter via the the
vacuum expectation value (VEV) of a SM singlet $S$. One simple
possibility\footnote{The $\mu$ parameter can also be generated in models
with no additional gauge groups, {\it i.e.} the next-to-minimal
supersymmetric standard model (NMSSM).  However, NMSSM models generically
possess discrete vacua and the tensions of the walls separating them are
too large to be cosmologically admissable \cite{nmssm}.} 
\cite{cveticlangacker} is to invoke an
additional nonanomalous $U(1)'$ gauge symmetry broken at the TeV 
scale, as expected in many string models. For suitable $U(1)'$ 
charges, the bare $\mu$ parameter is forbidden but the operator $h_s S 
H_u\cdot H_d$ is allowed, such that an effective $\mu$ term is 
generated after $S$ develops a VEV of order the electroweak/TeV scale 
(assuming the Yukawa coupling $h_s \sim O(1)$, as is well-motivated within 
semirealistic superstring models).
This framework is of particular interest because such extra $U(1)$ groups 
are often present in plausible extensions of the MSSM, and in fact are 
ubiquitous within many classes of four-dimensional superstring models.
Indeed, additional nonanomalous $U(1)$ gauge groups are present in 
virtually all known 4D string models with semirealistic features, such as 
gauge structure which includes $SU(3)_c\times SU(2)_L\times U(1)_Y$ (or a 
viable GUT extension) and particle content which includes the MSSM 
fields.\footnote{For example, many examples of such semirealistic models 
have been constructed within perturbative heterotic string theory
(see e.g. \cite{Quevedo:1996sv} for an overview).  An interesting class of 
constructions is the set of free fermionic models 
\cite{faraggistringmodels,chl,leontarisrizos}, in which a number of extra 
$U(1)$'s are always present at the string scale.  Whether or not all of 
these $U(1)$s 
persist to the TeV scale depends on the details of the vacuum 
restabilization procedure. Although there are cases in which only the MSSM 
gauge structure remains at low energy \cite{minimalstring}, typically one 
or more extra $U(1)$s persists to the electroweak scale 
\cite{faraggiphenomenology,chlphenom}. 
Additional $U(1)$s also are generic in supersymmetric braneworld models 
derived from Type II string orientifolds \cite{orientifold} (due at least 
in part to the $U(N)$ gauge groups associated with a stacks of D branes).  
Phenomenological analyses also indicate that typically extra $U(1)$s are 
present in the low energy theory and broken at the electroweak/TeV scale 
\cite{Cvetic:2002qa}.}

Within this class of models, the electroweak and $U(1)'$ symmetry 
breaking is driven by the soft SUSY breaking parameters, and hence the 
$Z'$ mass is expected to be of order a few TeV or less.  Such a 
$Z'$ should be easily observable at either present or forthcoming 
colliders. Indeed, the nonobservation to date of a $Z'$ puts 
interesting but stringent limits on the $Z'$ mass and mixing with 
the ordinary $Z$ both from direct searches at the Tevatron 
\cite{Abe:1997fd} and indirect tests from precision electroweak 
measurements \cite{indirect}. Although limits depend
on the details of the $Z'$ couplings, typically
$M_{Z'}>500-800$ GeV and the $Z-Z'$ mixing angle
$\alpha_{Z-Z'}\simlt O(10^{-3})$.\footnote{A potentially more
stringent limit on the $Z'$ mass arises from cosmology if the $U(1)'$ 
gauge symmetry forbids the standard implementation of the seesaw mechanism 
for neutrino masses.  In such scenarios, the right-handed neutrinos may be light,
and BBN constraints then require model-dependent limits
that in some cases are as strong as $M_{Z'}\simgt 4$ TeV \cite{Barger:2003zh}.} 
These models have been analyzed at tree 
level in \cite{Cvetic:1997ky,Langacker:1998tc,Langacker:1999hs}, where it 
was found that there are corners of parameter space in which an acceptable
$Z-Z'$ hierarchy can be achieved.  Further studies of a different class
of string-motivated $U(1)'$ models can be found in \cite{moreU1}.

As the phase of the $\mu$ parameter filters into the amplitudes for many
physical observables in the MSSM (and plays an important role in the 
Higgs sector at one-loop), it is worthwhile to analyze models which
solve the $\mu$ problem in the presence of explicit CP violation. In this
paper, we thus study the supersymmetric CP problem in $U(1)'$
models, focusing on the radiative corrections to the Higgs sector of the
$U(1)'$ model of \cite{Cvetic:1997ky} in the case that the soft
supersymmetry breaking parameters have general CP-violating phases
(radiative corrections in the CP-conserving case has been studied in
\cite{Amini:2002jp}).  We begin by classifying the reparameterization
invariant phase combinations and comment on the phenomenological
constraints on these phases from EDM bounds.  We then turn to the Higgs
sector, which conserves CP at tree level, but phases enter the Higgs 
potential through the stop mass-squared matrix at one-loop (just as in the 
MSSM).  The VEV's of the electroweak Higgs 
doublets $H_{u,d}$ and singlet $S$ are then determined by minimizing the 
loop-corrected Higgs potential. Within this framework, an effective $\mu$ 
parameter of the correct magnitude is generated which also has a phase 
governed by the phases of the soft SUSY breaking parameters.  We study the 
pattern of Higgs masses and mixings including the EDM and $Z'$ constraints,
and discuss the phenomenological implications for Higgs searches.

\section{The SUSY CP Problem in $U(1)'$ Models}

We study the class of $U(1)'$ models of \cite{Cvetic:1997ky}, in which the 
gauge group is extended to 
\begin{equation} 
G=\mbox{SU(3)}_{c}\times \mbox{SU(2)}_{L}\times
\mbox{U(1)}_{Y} \times \mbox{U(1)}',
\end{equation} 
with gauge couplings $g_3,g_2,g_Y,g_{Y'}$, respectively.
The matter content includes the MSSM superfields 
and a SM singlet $S$, which are all generically assumed to be charged
under the additional $U(1)'$ gauge symmetry.  Explicitly, the
particle content is: $\widehat{L}_i \sim (1,2,-1/2, Q_{L})$,
$\widehat{E}_i^{c} \sim (1, 1, 1, Q_{E})$, $\widehat{Q}_i \sim (3, 2, 1/6,
Q_{Q})$, $\widehat{U}_i^{c} \sim (\bar{3}, 1, -2/3, Q_{U})$,
$\widehat{D}_i^{c} \sim (\bar{3}, 1, 1/3, Q_{D})$, $\widehat{H}_{d} \sim
(1, 2, -1/2, Q_d)$, $\widehat{H}_{u} \sim (1, 2, 1/2, Q_u)$,
$\widehat{S}\sim (1, 1, 0, Q_{S})$, in which $i$ is the
family index.\footnote{Note that if the $U(1)'$ charges are family
nonuniversal they provide a tree-level source of FCNC. Phenomenological
bounds thus dictate that the charges of the first and second families
should be identical to avoid overproduction of FCNC without fine-tuning
\cite{Langacker:2000ju}.}

The superpotential includes a Yukawa coupling of the two electroweak Higgs
doublets $H_{u,d}$ to the singlet $S$, as well as a top quark Yukawa
coupling:
\begin{equation} 
W=h_{s}\widehat{S} \widehat{H}_{u}\cdot\widehat{H}_{d} + h_t
\widehat{U}_3^c\widehat{Q}_3\cdot\widehat{H}_u.  
\label{superpot} 
\end{equation} 
Gauge invariance of $W$ under $U(1)'$ requires that $Q_u+Q_d+ 
Q_{S} = 0$ and $Q_{Q_3}+Q_{U_3}+Q_u=0$. This choice of 
charges not only 
forbids the ``bare" $\mu$ parameter but also a K\"ahler potential 
coupling of the form $H_u H_d+h.c.$ required for the Giudice-Masiero 
mechanism \cite{giudice} (the K\"ahler potential is otherwise assumed to 
be of canonical form).\footnote{Other than these constraints, we prefer 
to leave the $U(1)'$ charges unspecified because our aim is not 
to construct a specific model.  In an explicit 
model there will be additional constraints on the $U(1)'$ charges (e.g 
from anomaly cancellation).  We also do not consider kinetic mixing in the 
analysis \cite{kinmix}.  However, even if kinetic mixing is absent at tree 
level will be generated through 1-loop RG running if ${\rm 
Tr}Q_YQ_{1'}\neq 0$ \cite{kinmix}.}

The form of (\ref{superpot}) is motivated by string models in which a
given Higgs doublet only has $O(1)$ Yukawa couplings to a single (third)
family.  We will consider the 
small $\langle H_u\rangle/\langle H_d\rangle\equiv \tan\beta$ regime 
only\footnote{Here low values of $\tan\beta$ such as 
$\tan\beta=1$ are allowed (this region is excluded in 
the MSSM).  The reason is that the Higgs bosons are generically 
heavier in $U(1)'$ models (as in the NMSSM and other models with extended 
Higgs sectors), and even at tree level the lightest Higgs 
boson can easily escape LEP bounds.}  such 
that the Yukawa couplings of the $b$ and $\tau$ can 
be safely neglected. The origin of the Yukawa couplings of the first and 
second generations of quarks and leptons is not addressed.  As we are 
primarily interested in the third family, we shall suppress the family 
index in what follows.

The soft supersymmetry breaking parameters include gaugino masses $M_a$ 
($a=1,1',2,3$), trilinear couplings $A_s$ and $A_t$, and soft 
mass-squared parameters 
$m_{\alpha}^2$: 
\begin{eqnarray}
\label{soft}
-{\cal L}_{soft}&=&(\sum_{a}M_a\lambda_a\lambda_a+
   A_s h_{s} S H_{u}\cdot H_{d}+ 
A_t h_t \widetilde{U}^c \widetilde{Q}\cdot H_u+h.c.)
  + m_{u}^{2}|H_{u}|^2 +  m_{d}^{2}|H_{d}|^2\nonumber\\
&+&m_{s}^{2}|S|^2+M_{\widetilde{Q}}^2|\widetilde{Q}|^2+ 
M_{\widetilde{U}}^2|\widetilde{U}|^2  
+ M_{\widetilde{D}}^2|\widetilde{D}|^2+ 
M_{\widetilde{E}}^2|\widetilde{E}|^2+M_{\widetilde{L}}^2|\widetilde{L}|^2.
\end{eqnarray}
These soft SUSY breaking parameters are generically nonuniversal at low
energies. We do not address the origin of these low energy parameters via
RG evolution from high energy boundary conditions in this paper.

The gaugino masses $M_a$ and soft trilinear couplings $A_{s,t}$ of 
(\ref{soft}) can be complex; if so, they can provide sources of CP 
violation (without loss of generality, the Yukawa couplings $h_{s,t}$ can 
be assumed to be real). However, not all of these phases are physical, 
just as the case in the MSSM.  Let us first consider the MSSM.  The 
reparameterization invariant combinations of phases in the MSSM are easily 
determined by forming invariants with respect to the global U(1)$_{PQ}$ 
and U(1)$_R$ symmetries present in the limit that the soft breaking 
parameters and the $\mu$ term are set to zero \cite{Dimopoulos:1995kn}; 
for reference, the $U(1)_{R,PQ}$ charge assignments are presented in 
Table~\ref{table1}. A convenient basis of the resulting reparameterization 
invariant phases thus is $\theta_{f}=\phi_{\mu}+\phi_{A_f}-\phi_{b}$ and 
$\theta_{a}=\phi_{\mu}+\phi_{M_a}-\phi_{b}$, which enter the mass 
matrices of the sfermions and the gauginos/Higgsinos, respectively. 
An analysis of the MSSM tree level Higgs sector also 
suggests it is useful to exploit U(1)$_{PQ}$ to set 
$\phi_b=0$ ($\phi_b$ is then dropped from the invariants above), in which 
case the Higgs VEVs are real.   

\begin{table}[tbp] 
\begin{center} 
\label{table1}
\begin{tabular}{|c||c|c|c|c|c|c|c|c|c|c|c|} 
\hline 
Field & $\widehat{Q}$ & $\widehat{U}^c$ & $\widehat{D}^c$ & 
$\widehat{H}_u$ & $\widehat{H}_d$ &  $\lambda_a$ & $M_a$ & $\mu$&$b$&  
$A_{f}$ & $m_{\alpha}^{2}$ \\ \hline
U(1)$_R$ & 1 & 1 & 1 & 0 & 0 & 1 & $-$2 & 2 & 0&$-$2  &0\\ \hline 
U(1)$_{PQ}$ & 0 & 0 &0 & 1 & 1 & 0& 0 & $-$2& $-$2& 0&0\\ \hline 
\end{tabular} 
\end{center} 
\caption{\footnotesize The $U(1)_{R,PQ}$ charge assignments for the MSSM 
fields and spurions.}
\end{table}

Performing the same exercise in the $U(1)'$ framework, one immediately 
notices that the U(1)$_{PQ}$ symmetry of the MSSM is embedded within the 
$U(1)'$ gauge symmetry.  However, a nontrivial U(1)$_R$ symmetry 
remains;  the $U(1)_R$ charges of the superfields and the associated 
spurion charges of the soft parameters are presented in Table 
\ref{table2}.  The reparameterization
invariant phase combinations are therefore
$\theta_{f\,f'}=\phi_{A_f}-\phi_{A_{f'\neq f}}$,
$\theta_{a\,f}=\phi_{M_a}-\phi_{A_f}$, and
$\theta_{a\,b}=\phi_{M_a}-\phi_{M_{b\neq a}}$, of which only two are
linearly independent (e.g. $\phi_{a\,b}=\phi_{a\,f}-\phi_{b\,f}$).
We will see that (in analogy to the MSSM) the tree-level Higgs sector 
suggests it is convenient to measure all phases with 
respect to the phase of $A_s$.  (In fact, one can go further and exploit 
the $U(1)_R$ symmetry to set $\phi_{A_s}=0$, although we prefer not to do 
that in this paper.) A basis of reparameterization invariant phase 
combinations can then be chosen as
\begin{eqnarray}
\label{u1pphases}
\theta_{f\,s}&=&\phi_{A_f}-\phi_{A_s}\nonumber \\ 
\theta_{a\,s}&=&\phi_{M_a}-\phi_{A_s}.      
\end{eqnarray}

\begin{table}[tbp] 
\begin{center} 
\label{table2}
\begin{tabular}{|c||c|c|c|c|c|c|c|c|c|c|} 
\hline 
Field & $\widehat{Q}$ & $\widehat{U}^c$ & $\widehat{D}^c$ & 
$\widehat{H}_u$ & $\widehat{H}_d$ & $\widehat{S}$ & $\lambda_a$ & $M_a$ &  
$A_{f}$ & $m_{\alpha}^{2}$ \\ \hline
U(1)$_R$ & 1 & 1 & 1 & 0 & 0 & 2 & 1 & -2 & - 2 & 0\\ \hline 
\end{tabular} 
\end{center} 
\caption{\footnotesize The U(1)$_R$ charge assignments for the fields
and spurions in the $U(1)'$ framework. Note that $\widehat{S}$ 
(whose VEV induces an effective $\mu$ parameter) 
has nonzero $R$ charge.}
\end{table}
To see this more explicitly and to lay the foundation for our analysis of 
the Higgs sector including one-loop radiative corrections, let us now 
review the tree-level Higgs potential analyzed in \cite{Cvetic:1997ky}. 
Gauge symmetry breaking is driven by the VEVs of the electroweak Higgs 
doublets $H_u$, $H_d$
\begin{eqnarray}
H_u=\left(\begin{array}{c c} 
H_u^+\\H_u^0\end{array}\right),\;\;
H_d=\left(\begin{array}{c c} H_{d}^0\\H_d^-\end{array}\right),
\end{eqnarray}
and the singlet $S$. The tree level Higgs 
potential is a sum of F terms, D terms, and soft supersymmetry breaking 
terms:
\begin{equation}
\label{treepot}
V_{tree}=V_F+V_D+V_{soft},
\end{equation}
in which
\begin{eqnarray}
\label{vf}
V_F&=& |h_{s}|^2\left[ |H_u\cdot H_d|^2+ |S|^2 ( 
|H_u|^2+|H_d|^2)\right],
\end{eqnarray}
\begin{eqnarray}
\label{vd}
V_D&=&\frac{G^2}{8}\left(|H_u|^2-|H_d|^2\right)^2+
   \frac{g_{2}^2}{2}(|H_u|^2|H_d|^2-|H_u \cdot H_d|^2)\nonumber \\
&+&\frac{g_{Y'}^2}{2}\left(Q_u|H_u|^2+Q_d|H_d|^2+
   Q_S|S|^2\right)^2,
\end{eqnarray}
\begin{equation}
V_{soft}=m_{u}^{2}|H_u|^2+
  m_{d}^{2}|H_d|^2+
  m_s^{2}|S|^2 +(A_sh_sSH_u\cdot H_d+h.c.).
\end{equation}
where $G^{2}=g_2^{2}+g_Y^{2}$ and $g_Y=\sqrt{3/5} g_1$, $g_1$ is the
GUT normalized hypercharge coupling.

At the minimum of the potential, the Higgs fields are
expanded\footnote{As discussed in \cite{Cvetic:1997ky}, gauge rotations
can be used to set $\langle H_u^+\rangle=0$.  However, $\langle H_d^-
\rangle=0$ is not automatic and imposes constraints on the parameter space
of the model. Indeed, $\langle H_d^-\rangle=0$ implies that the physical 
charged Higgs is nontachyonic ($M_{H^{\pm}}^{2} >0$).} as follows:
\begin{eqnarray}
\label{higgsexp}
\langle H_u \rangle &=& \frac{1}{\sqrt{2}}\left(\begin{array}{c} \sqrt{2}
H_u^{+}\\ v_u + \phi_u + i \varphi_u\end{array}\right)  \:,\:\:\:\:
\langle H_d \rangle = \frac{1}{\sqrt{2}}\left(\begin{array}{c} v_d +
\phi_d + i \varphi_d\\ \sqrt{2} H_d^{-}\end{array}\right)\nonumber\\   
\langle S \rangle &=& \frac{1}{\sqrt{2}} e^{i\theta} \left(v_s + \phi_s +
i \varphi_s\right),
\end{eqnarray}
in which $v^2\equiv v_u^2+v_d^2=(246\,{\rm GeV})^2$.  In the above, a
phase shift $e^{i\theta}$ has been attached to $\langle S \rangle$.  Since 
gauge invariance dictates that only the phase of the combination
$SH_u\cdot H_d$ enters the potential, we can assume that the VEVs of
$H_{u,d}$ are real and attach a phase only to $S$ without loss of
generality (this choice is also consistent with our assignment of $U(1)_R$
charges in Table \ref{table1}).  The effective $\mu$ parameter 
is generated by the singlet VEV $\langle S \rangle$:
\begin{eqnarray} 
\label{mupar} 
\mu_{eff} 
\equiv \frac{h_s v_s}{\sqrt{2}}\ e^{i \theta}. 
\end{eqnarray} 
The only complex parameter which enters the Higgs potential at tree level
is $A_s$.  However, the global phases of the Higgs fields (more precisely,
of the combination $S H_u\cdot H_d$) can always be chosen to absorb the
phase $\phi_{A_s}$ of $A_s$ by performing a $U(1)_R$ rotation on the
fields, such that $A_s$ and the VEV's can 
all be taken to be real without loss of generality \cite{Cvetic:1997ky}. 
To state this another way, the minimization 
conditions with respect to the CP odd directions $\varphi_{1,2,s}$  
all lead to the condition
\begin{equation}
\label{treetadpole}
\sin(\theta+\phi_{A_s})=0,
\end{equation}
such that $\theta=-\phi_{A_s}$ at tree level.  With this condition, the
Higgs sector is CP conserving.  The Higgs mass eigenstates 
thus have definite CP quantum numbers, with three CP even
Higgs bosons $H_{i=1,2,3}$ and one CP odd Higgs boson $A^0$, as well as a 
charged Higgs pair $H^{\pm}$.  Expressions for their masses at tree level
and a discussion of the associated Higgs phenomenology can be found in 
\cite{Cvetic:1997ky}.

Although the Higgs sector conserves CP at tree level whether or not the
soft SUSY breaking parameters are complex, this is generically not true 
for other sectors of the theory and care must be taken in the 
phenomenological analysis (e.g. for the EDM bounds) if there are 
nontrivial CP-violating phases in the soft terms even if the Higgs sector 
is only analyzed at tree level.  Clearly, this is due to the fact that the 
phases which enter the mass matrices of the sfermion and the 
gaugino/higgsino sectors involve the phase of the singlet VEV $\theta$ 
({\it i.e.}, the phase of the effective $\mu$ parameter $\mu_{eff}$) as 
well as the phases of the $A$ terms and the gaugino masses.  For example, 
the reparameterization invariant phase combination which enters the 
chargino mass matrix within this class of $U(1)'$ models is
\begin{equation} 
\theta_{\widetilde{\chi}^{\pm}}=\theta+\phi_{M_2}= 
\phi_{M_2}-\phi_{A_s}+\ldots = \theta_{2\,s}+\ldots,
\end{equation}
in which the terms represented by $(+\ldots)$ are higher-loop 
contributions. As previously mentioned, this phase is strongly 
constrained by EDM experimental bounds (although the precise constraints 
can depend in detail on the other parameters of the model).  More 
generally, the statement of (flavor-independent) SUSY CP  
violation within $U(1)'$ models is that if any of the phases 
$\theta_{f\,s}$ and $\theta_{a\,s}$ defined in (\ref{u1pphases}) are 
nonzero, they can lead to 
CP-violating effects which may be in conflict with experiment and 
must be checked.  This is in direct analogy to the statement of 
flavor-independent SUSY CP violation in the MSSM. However, as $\mu$ is 
dynamically generated within $U(1)'$ models, its phase $\phi_{\mu}$ 
is now a function of the phases of the other soft breaking 
parameters rather than an independent quantity.
 
Returning now to the question of CP violation in the Higgs sector, 
(\ref{treetadpole}) suggests that it is natural to consider the 
combination of phases
\begin{equation}
\label{thetaprdef}
\overline{\theta}\equiv \theta+\phi_{A_s} 
\end{equation} 
as the parameter which governs CP violation in the Higgs sector.  Note
that $\overline{\theta}$ is a reparameterization invariant quantity, while
$\theta$ is not ($\theta=\overline{\theta}$ in the basis in which
$\phi_{A_s}=0$).  While $\overline{\theta}=0$ at tree level, it acquires a 
nonzero value at one-loop if the sfermion and gaugino/higgsino mass 
matrices have nontrivial phases.  This calculation is outlined in the next 
section.

\section{Higgs Sector CP Violation}
Previously we discussed the SUSY CP problem within $U(1)'$ models,
and reviewed the tree level Higgs sector (the patterns of gauge symmetry
breaking which led to an acceptable $Z-Z'$ hierarchy were analyzed
at tree level in \cite{Cvetic:1997ky}).  In what follows, we will compute
the one-loop radiative corrections to the Higgs sector of this class of
$Z'$ models within a general framework including nontrivial CP
violation (radiative corrections in the CP-conserving case were previously
presented in \cite{Amini:2002jp}).

\subsection{Radiative Corrections to the Higgs Potential}
The effective potential approach provides an elegant way 
of determining the true vacuum state of a spontaneously broken
gauge theory. The potential has the form
\begin{equation}
V=V_{tree} + \Delta V+\ldots,
\end{equation}
where $V_{tree}$ is defined in (\ref{treepot}), and the one-loop
contribution $\Delta V$ has the Coleman-Weinberg form
\begin{equation}
\Delta V = \frac{1}{64 \pi^{2}}\left \{ \mbox{Str}
{\cal{M}}^{4}(H_u,H_d,S)\left(\ln{\frac{{\cal{M}}^{2}(H_u,H_d,S)}{Q^{2}}}-
\frac{3}{2}\right)\right \}.
\end{equation}
in the mass-independent renormalization scheme
$\overline{\mbox{DR}}$.\footnote{See Martin's paper in \cite{effpot} for a 
detailed discussion of the regularization and renormalization scheme 
dependence of the effective potential.} In the above, $\mbox{Str}\equiv 
\sum_{J} (-1)^{2 J +1} (2 J + 1)$ is the usual supertrace, $Q$ is the
renormalization scale, and ${\cal{M}}$ represents the Higgs
field-dependent mass matrices of the particles and sparticles of the
theory.\footnote{While the complete effective potential is scale
invariant, it is scale dependent when truncated to any finite loop order 
in perturbation theory due to the renormalization of the 
parameters and the Higgs wavefunctions. In the MSSM, most of the 
scale-dependent terms can be collected in the pseudoscalar mass, which 
itself can be regarded as a free parameter of the theory. The remaining 
$Q^{2}$-dependence arises from the D term contributions generated by 
wavefunction renormalization, such that in the limit in which $g_2=g_Y=0$ 
all of the scale dependence can be absorbed into the pseudoscalar mass. 
For the $U(1)'$ models, the scale dependence can be absorbed into the 
pseudoscalar mass only if the D term contributions vanish {\it and} the 
superpotential parameter $h_s=0$, because the potential also includes 
quartic Higgs couplings which arise from F terms. These properties are 
manifest in the Higgs mass-squared matrix presented below.}

Here we will include only the dominant terms due to top and scalar top 
quark loops:
\begin{eqnarray}
\label{colemanweinberg}
\Delta V = \frac{6}{64 \pi^2}\ \left\{ \sum_{k=1,2}
\left(m_{\widetilde{t}_{k}}^{2}\right)^{2} \left[
\ln\left(\frac{m_{\widetilde{t}_{k}}^{2}}{Q^2}\right) - \frac{3}{2}\right]
- 2 \left(m_t^2\right)^{2} \left[ \ln\left(\frac{m_t^{2}}{Q^2}\right) -   
\frac{3}{2}\right]\right\}
\end{eqnarray}
in which the masses depend explicitly on the Higgs field components
(note that $\Delta V$ naturally vanishes in the limit of exact SUSY).  The 
top mass-squared is given by $m_t^{2}= h_t^{2} |H_u|^2$, and the stop 
masses-squared are obtained by diagonalizing the mass-squared matrix
\begin{eqnarray}
\widetilde{M}^{2}=\left(\begin{array}{cc} M_{LL}^{2} & M_{LR}^{2}\\
{M^{2}_{LR}}^{\dagger} & M_{RR}^{2}\end{array}\right)
\end{eqnarray}
via the unitary matrix $S_t$ as $S_t^{\dagger} \widetilde{M}^{2} S_t =
{\mbox{diag}}\left(m_{\widetilde{t}_{1}}^{2},
m_{\widetilde{t}_{2}}^{2}\right)$. The entries of $\widetilde{M}^{2}$ are
given by
\begin{eqnarray}
M_{LL}^{2}&=&M_{\widetilde{Q}}^{2}+h_t^{2} |H_u|^2
-\frac{1}{4}\left(g_2^{2}-\frac{g_Y^{2}}{3}\right)
\left(|H_u|^2 - |H_d|^2 \right)\nonumber\\&+&
g_{Y'}^{2} Q_Q \left(Q_u |H_u|^2 + Q_d |H_d|^2
+ Q_s |S|^2 \right)\\
M_{RR}^{2}&=&M_{\widetilde{U}}^{2}+h_t^{2} |H_u|^2
-\frac{1}{3}g_Y^{2} \left(|H_u|^2 - |H_d|^2 \right)\nonumber\\
&+& g_{Y'}^{2} Q_U \left(Q_u |H_u|^2+ Q_d |H_d|^2 + 
Q_s|S|^2 \right)\\
M_{LR}^{2}&=&h_t \left(A^*_t H_u^{0*}-h_s S 
H_d^{0}\right),\: 
\end{eqnarray}
in which we have emphasized the fact that the LR entry depends 
only on the neutral components of the Higgs fields. As the stop LR 
mixing can be complex, the term $h_t h_s A_t S {H_u^{0}} H_d^{0} +
\mbox{h.c.}$ present in $\left|M_{LR}^{2}\right|^{2}$ can 
provide a source of CP violation in the Higgs sector.  From the discussion 
of the previous section, we can infer that this source is the phase 
$\theta_{t\,s}\equiv \phi_{A_t}-\phi_{A_s}$.

The vacuum state is characterized by the vanishing of all tadpoles and
positivity of the resulting Higgs boson masses. Recalling the 
expressions for the Higgs fields in (\ref{higgsexp}), the vanishing of 
tadpoles for $V$ along the CP even directions $\phi_{u,d,s}$ enables the
soft masses $m_{u,d,s}^{2}$ to be expressed in terms of the other 
parameters of the potential:
\begin{eqnarray}
\label{massesu}
m_u^{2}&=& M_{0}^{2} \cos^{2} \beta - \lambda_u v_u^{2} - \frac{1}{2}
\left( \lambda_{u d} v_d^{2} + \lambda_{u s} v_s^{2}\right)
        - \frac{1}{v_u} \left(\frac{\partial \Delta V}{\partial
\phi_u}\right)_{0}\\
\label{massesd}
m_d^{2}&=& {M}_{0}^{2} \sin^{2} \beta - \lambda_d v_d^{2} - \frac{1}{2}
\left( \lambda_{u d} v_u^{2} + \lambda_{d s} v_s^{2}\right)
        - \frac{1}{v_d} \left(\frac{\partial \Delta V}{\partial
\phi_d}\right)_{0}\\
\label{massess}
m_s^{2}&=& {M}_{0}^{2} \cot^{2} \alpha - \lambda_s v_s^{2} - \frac{1}{2}
\left( \lambda_{u s} v_u^{2} + \lambda_{d s} v_d^{2}\right)- \frac{1}{v_s} 
\left(\frac{\partial \Delta V}{\partial \phi_s}\right)_{0},
\end{eqnarray}  
in which the subscript 0 indicates that the derivatives of $\Delta V$ are 
to be evaluated at $\phi_i=0$ and $\varphi_i=0$. Here the various $\lambda$
coefficients represent the quartic couplings in the potential
\begin{eqnarray}
\lambda_{u,d}&=&\frac{1}{8} G^{2}+ \frac{1}{2} Q_{u,d}^{2}
g_{Y'}^{2}\:\:\:,\:\:\:\:
\lambda_s=\frac{1}{2} Q_s^{2}\ g_{Y'}^{2}
\:\:\:,\:\:\:\:
%\widetilde{\lambda}_{u d}=\frac{1}{2} g_2^{2}- h_s^{2}
\nonumber\\
\lambda_{u d}&=& - \frac{1}{4} {G^{2}} + Q_u Q_d\
g_{Y'}^{2} + h_s^{2}\:\:\:,\:\:\:\: \lambda_{us, ds}= Q_s
Q_{u,d}\
g_{Y'}^{2} + h_s^{2}\: .
\end{eqnarray} 
The Higgs soft masses (\ref{massesu}) are written in terms
of two angle parameters: (i) $\tan\beta$, which measures the hierarchy of 
the Higgs doublet VEVs, and (ii) $\cot\alpha \equiv (v \sin\beta 
\cos\beta)/v_s$, which is an indication of the splitting 
between the $U(1)'$ and electroweak breaking scales. For 
convenience, we have also introduced the mass parameter
\begin{eqnarray}
{M}_{0}^{2}= \frac{h_s |A_s| v_s \cos \overline{\theta}}{\sqrt{2} \sin\beta
\cos\beta},
\end{eqnarray}
which corresponds to the mass parameter of the CP odd Higgs boson of
the MSSM after using the definition of the effective $\mu$ parameter in
(\ref{mupar}).

While the vanishing of the tadpoles along the CP odd directions 
$\varphi_{u,d,s}$ led to (\ref{treetadpole}) at tree level, once the loop 
corrections 
are included they lead to the following conditions:
\begin{eqnarray}
\label{pseuu}
{M}_{0}^{2} \sin\beta \cos\beta \tan \overline{\theta} &=& \frac{1}{v_d}
\left(\frac{\partial \Delta V}{\partial \varphi_u}\right)_{0}\\
\label{pseud}
{M}_{0}^{2} \sin\beta \cos\beta \tan \overline{\theta} &=& \frac{1}{v_u}
\left(\frac{\partial \Delta V}{\partial \varphi_d}\right)_{0}\\
\label{pseus}
{M}_{0}^{2} \cot\alpha \tan \overline{\theta} &=& \frac{1}{v_s} 
\left(\frac{\partial \Delta V}{\partial \varphi_s}\right)_{0},
\end{eqnarray}
demonstrating explicitly that the phase 
$\overline{\theta}=\theta+\phi_{A_s}$ associated with the phase $\theta$ of 
the singlet VEV is indeed a radiatively induced quantity. Indeed, the 
derivatives of $\Delta V$ with respect to $\varphi_{u,d,s}$ are 
nonvanishing provided that there is a nontrivial phase difference 
between $A_t$ and $A_s$ ({\it i.e.}, if $\theta_{t\,s}\neq 0$).  
In fact, (\ref{pseuu})--(\ref{pseus}) all lead to the same relation for 
$\overline{\theta}$:
\begin{eqnarray}
\label{theta}
\sin\overline{\theta}= - \beta_{h_t}\ \frac{|A_t|}{|A_s|} \sin \theta_{t\, s}\ 
{\cal{F}}(Q^2, m_{\widetilde{t}_{1}}^{2},
m_{\widetilde{t}_{2}}^{2})
\end{eqnarray}
in which $\beta_{h_t}=3 h_t^2/(32 \pi^{2})$ is the beta 
function for the top Yukawa coupling, and the loop function 
\begin{eqnarray}
{\cal{F}}(Q^2, m_{\widetilde{t}_{1}}^{2},
m_{\widetilde{t}_{2}}^{2})= -2 + \ln\left(
\frac{m_{\widetilde{t}_{1}}^{2} m_{\widetilde{t}_{2}}^{2}}{Q^4}\right) +
\frac{m_{\widetilde{t}_{1}}^{2} +
m_{\widetilde{t}_{2}}^{2}}{m_{\widetilde{t}_{1}}^{2}-
m_{\widetilde{t}_{2}}^{2}}\
\ln\left(\frac{m_{\widetilde{t}_{1}}^{2}}{m_{\widetilde{t}_{2}}^{2}}\right)
\end{eqnarray}
depends explicitly on the renormalization scale. In the above, 
$\theta_{t\, s}=\phi_{A_t}+\theta=\phi_{A_t} - \phi_{A_s}$ up to
one loop accuracy determined by (\ref{theta}).

\subsection{The Higgs Mass Calculation}

We now turn to the Higgs mass calculation at one-loop in the presence of 
CP violation in the stop LR mixing.  The mass-squared matrix of the Higgs 
scalars is 
\begin{eqnarray}
{\cal{M}}^{2}_{i j}=\left(\frac{\partial^{2}}{\partial \Phi_i \partial
\Phi_j} V\right)_{0},
\end{eqnarray}
subject to the minimization constraints 
Eqs.~\ref{massesu}-\ref{massess} and (\ref{theta}). In the above, 
$\Phi_i=(\phi_i,\varphi_i)$. Clearly, two linearly independent 
combinations of the pseudoscalar components $\varphi_{u,d,s}$ are the 
Goldstone bosons $G_Z$ and $G_{Z'}$, which are eaten by the $Z$ and 
$Z'$ gauge bosons when they acquire their masses. These two modes 
are given by
\begin{eqnarray}
G_Z=-\sin\beta \varphi_u + \cos\beta \varphi_d\:,\:\:\:
G_{Z'}=\cos\beta \cos\alpha \varphi_u + \sin\beta \cos\alpha
\varphi_d -\sin\alpha \varphi_s,
\end{eqnarray}
and hence the orthogonal combination 
\begin{eqnarray}
A=\cos\beta \sin\alpha \varphi_u + \sin\beta \sin\alpha \varphi_d
+\cos\alpha \varphi_s
\end{eqnarray}
is the physical pseudoscalar Higgs boson in the CP-conserving limit. In 
the decoupling limit, $v_s\gg
v$, $\sin\alpha\rightarrow 1$ and $\cos\alpha\rightarrow 0$, in which case 
$G_Z$ and $A$ reduce to their MSSM expressions. In the basis of  
scalars ${\cal{B}}=\left\{\phi_u, \phi_d, \phi_s, A\right\}$, the Higgs 
mass-squared matrix ${\cal{M}}^2$ takes the form
\begin{eqnarray}
\left(\begin{array}{cccc}
\label{higgsmassmat}
M_{u u}^{2} + M_A^{2} \cos^2\beta & M_{u d}^{2} - M_A^{2} \sin\beta
\cos\beta & M_{u s}^{2}- M_{A}^{2} \cot\alpha \cos\beta
 & M_{u A}^{2} \sin\theta_{t\, s} \\\\ M_{u d}^{2} - M_A^{2} \sin\beta \cos\beta &
M_{d d}^{2} + M_A^{2} \sin^{2}\beta & M_{d s}^{2}- M_{A}^{2} \cot\alpha 
\sin\beta & M_{d A}^{2} \sin\theta_{t\, s} \\\\ M_{u s}^{2}- M_{A}^{2} \cot\alpha
\cos\beta & M_{d s}^{2}- M_{A}^{2} \cot\alpha \sin\beta & M_{s s}^{2} +
M_A^{2} \cot^{2}\alpha & M_{s A}^{2} \sin\theta_{t\, s} \\\\ M_{u A}^{2}
\sin\theta_{t\, s} & M_{d A}^{2} \sin\theta_{t\, s} & M_{s A}^{2}\sin\theta_{t\, s} &
M_{P}^{2}\end{array}\right),
\end{eqnarray}
in which our notation explicitly demonstrates that all of the entries
${\cal{M}}^2_{i A}$ ($i=u,d,s$) identically vanish in the CP-conserving
limit $\theta_{t\, s}\rightarrow \phi_0$. In the above,
\begin{eqnarray}
M_A^{2}={M}_0^{2}\left(1 + \beta_{h_t} \frac{|A_t|}{|A_s|}\ \frac{\cos
\theta_{t\, s}}{\cos\overline{\theta}}\ {\cal{F}}\right),
\end{eqnarray}
which depends explicitly on the renormalization scale, and
\begin{eqnarray}
\label{pseumass}
M_{P}^{2}=\frac{M_A^2}{\sin^2\alpha} + 4 \beta_{h_t} \frac{ m_t^{2}
|\mu_{eff}|^{2} |A_t|^{2}}
{\left(m_{\widetilde{t}_{1}}^{2}-m_{\widetilde{t}_{2}}^{2}\right)^{2}}\
\frac{\sin^{2}\theta_{t\, s}}{\sin^{2}\alpha \sin^2\beta} {\cal{G}}
\end{eqnarray}
is the one-loop pseudoscalar mass in the CP-conserving limit. The
loop function ${\cal{G}}$ is independent of the renormalization scale and 
has the functional form 
\begin{eqnarray}
{\cal{G}}\left(m_{\widetilde{t}_{1}}^{2},
m_{\widetilde{t}_{2}}^{2}\right)= 2 - \frac{m_{\widetilde{t}_{1}}^{2} +
m_{\widetilde{t}_{2}}^{2}}{m_{\widetilde{t}_{1}}^{2}-
m_{\widetilde{t}_{2}}^{2}}\
\ln\left(\frac{m_{\widetilde{t}_{1}}^{2}}{m_{\widetilde{t}_{2}}^{2}}
\right)\:.
\end{eqnarray}
We now turn to the mass parameters $M_{i j}^{2}$ ($i,j=u,d,s$) 
which appear in ${\cal{M}}^2$. These entries may be represented as
\begin{eqnarray}
\label{corr}
M_{i j}^{2} &=& v_i v_j \Bigg\{ \overline{\lambda}_{i j} + \frac{3}{(4 
\pi)^{2}} \Bigg[ \frac{\left(\rho_i \widetilde{m}_j^{2} +
\widetilde{m}_i^{2}
\rho_j\right)}{m_{\widetilde{t}_{1}}^{2}+m_{\widetilde{t}_{2}}^{2}}    
(2-{\cal{G}}) + \left( \rho_i \rho_j + \zeta_i \zeta_j + \delta_{i d}
\delta_{j s} \frac{h_t^2 h_s^2}{4}\right){\cal{F}}\nonumber\\
&+&\left(\rho_i \rho_j + \frac{\widetilde{m}_i^{2} \widetilde{m}_j^{2}}
{\left(m_{\widetilde{t}_{1}}^{2}-m_{\widetilde{t}_{2}}^{2}\right)^{2}}\right)
{\cal{G}} - \delta_{iu}\delta_{j u} h_t^{4}
\ln\left\{\frac{m_t^{4}}{Q^{4}}\right\}\Bigg]\Bigg\},
\end{eqnarray}
in which $\overline{\lambda}_{i
j}=\lambda_{i j}$ for $i\neq j$, $\overline{\lambda}_{i j}= 2
\lambda_{i}$ for $i=j$. For notational purposes we have also introduced 
the dimensionless quantities
\begin{eqnarray}
\rho_u=h_t^2-\lambda_u\:\:,\:\:\:  \rho_d=(h_s^2 - \lambda_{u
d})/2\:\:,\:\:\: \rho_s=(h_s^2-\lambda_{u s})/2
\end{eqnarray}  
as well as the dimensionful ones,
\begin{eqnarray}
\widetilde{m}_u^{2}&=&\zeta_u \delta + h_t^2 |A_t| \left(|A_t|-|\mu_{eff}|
\cot\beta \cos\theta_{t\, s}\right)\\ \widetilde{m}_d^{2}&=&\zeta_d
\delta + h_t^2 |\mu_{eff}| \left(|\mu_{eff}|-|A_t| \tan\beta
\cos\theta_{t\, s}\right)\\ \widetilde{m}_s^{2} &=& \zeta_s \delta +
\frac{v_d^2}{v_s^2} h_t^2 |\mu_{eff}| \left(|\mu_{eff}|-|A_t| \tan\beta
\cos\theta_{t\, s}\right),
\end{eqnarray}
with $\delta=M_{\widetilde{Q}}^{2}- M_{\widetilde{U}}^{2}+\zeta_u v_u^2 +
\zeta_d v_d^2 + \zeta_s v_s^2$. The new dimensionless couplings appearing 
here are pure D term contributions
\begin{eqnarray}
\zeta_u &=& - \frac{1}{8} (g_2^2-\frac{5}{3}\, g_Y^2) + \frac{1}{2} 
(Q_Q-Q_U) Q_u g_{Y'}^{2}\\
\zeta_d &=& \frac{1}{8} (g_2^2- 
\frac{5}{3}\, g_Y^2) 
+\frac{1}{2} (Q_Q-Q_U) Q_d g_{Y'}^{2}\\
\zeta_s&=&-(\zeta_u + \zeta_d).
\end{eqnarray}
Finally, the scalar-pseudoscalar 
mixing entries $M_{i A}^{2}$ ($i=u,d,s$), which exist only if there are 
sources of CP violation in the Lagrangian (as has been made explicit in 
${\cal{M}}^2$ by factoring out $\sin\theta_{t\, s}$), are given by 
\begin{eqnarray}
M_{i A}^{2}= 2 \beta_{h_t}\ \frac{v v_i}{\sin\alpha}\ \frac{|\mu_{eff}|
|A_t|}{m_{\widetilde{t}_{1}}^{2}-m_{\widetilde{t}_{2}}^{2}} \left[ 2
\rho_i \frac{m_{\widetilde{t}_{1}}^{2}-m_{\widetilde{t}_{2}}^{2}}
{m_{\widetilde{t}_{1}}^{2}+m_{\widetilde{t}_{2}}^{2}}
+\left(\frac{\widetilde{m}_i^{2}}{m_{\widetilde{t}_{1}}^{2}-
m_{\widetilde{t}_{2}}^{2}} - \rho_i\frac{m_{\widetilde{t}_{1}}^{2}-
m_{\widetilde{t}_{2}}^{2}}
{m_{\widetilde{t}_{1}}^{2}+m_{\widetilde{t}_{2}}^{2}}\right)
{\cal{G}}\right],
\end{eqnarray}
and are scale independent. These results agree with the tree level 
computations of \cite{Cvetic:1997ky}. After identifying (\ref{mupar}) with 
the $|\mu_{eff}|$ parameter of the MSSM, the doublet sector of the 
mass-squared matrix agrees with that of the MSSM \cite{pilaftsis}. 
Finally, the results also agree with those of \cite{Amini:2002jp} in the 
CP-conserving limit ($\sin\theta_{t\, s}=0$).

As previously stated, there are three 
CP even and one CP odd Higgs boson in the CP-conserving limit.  The mass 
of the CP odd Higgs boson $A$ is given in (\ref{pseumass}), while the 
masses of the CP even scalars arise from the diagonalization of the upper 
$3\times 3$ subblock of (\ref{higgsmassmat}). The masses and mixings then 
differ from their tree level values by the inclusion of radiative effects.
In this limit, the only source of CP violation is the CKM matrix 
and one easily evades constraints from the absence of permanent
EDMs for leptons and hadrons. The lightest
Higgs boson has a larger mass than $M_Z$ even at tree level,
and the radiative effects modify it sizeably \cite{Amini:2002jp}. Once the
radiative corrections are included a direct comparison with experimental 
results is possible.  In principle, one can constrain certain portions of the 
parameter space using the post-LEP indications for a light scalar with 
mass $\simgt 114$ GeV.

In the presence of CP violation, there are four scalar bosons with no
definite CP quantum number. This results from the mixing between the CP
even scalars $\phi_{u,d,s}$ with the CP odd scalar $A$ via the entries
$M_{i A}^{2} \sin\theta_{t\, s}$ in (\ref{higgsmassmat}). The main impact
of the CP breaking Higgs mixings on the collider phenomenlogy comes via
the generation of novel couplings for Higgs bosons which eventually modify
the event rates and asymmetries. Indeed, a given Higgs boson can couple to
both scalar and pseudoscalar fermion densities depending on the strength
of CP violation \cite{pilaftsis}. Moreover, the coupling of the lightest
Higgs to $Z$ bosons can be significantly suppressed, avoiding the existing
bounds from the LEP data \cite{CPX,CPXX}. The CP-violating entries of
${\cal{M}}^{2}$ grow with $|\mu_{eff}A_t|$ as in the MSSM. The
mass-squared matrix is diagonalized by a $4\times 4$ orthonormal matrix
${\cal{R}}$
\begin{eqnarray}
{\cal{R}} \cdot M_{h}^{2}\cdot {\cal{R}}^{T} = 
\mbox{diag.}\left(M_{H_1}^{2},
M_{H_2}^{2}, M_{H_3}^{2}, M_{H_4}^{2}\right).
\end{eqnarray}
To avoid discontinuities in the eigenvalues it is convenient to
adopt an ordering: $M_{H_1}<M_{H_2}<M_{H_3}<M_{H_4}$. 
The mass eigenstates $H_i$ can then be expressed as
\begin{eqnarray}
H_i = {\cal{R}}_{i u} \phi_u + {\cal{R}}_{i d} \phi_d + {\cal{R}}_{i s}
\phi_s+ {\cal{R}}_{i A} A
\end{eqnarray}
in which e.g. $\left|{\cal{R}}_{i A}\right|^{2}$ is a measure
of the CP odd composition of $H_i$. The elements of ${\cal{R}}$ determine 
the couplings of Higgs bosons to the MSSM fermions, scalars, and gauge 
bosons.

\subsection{Comparison with MSSM}
Before turning to the numerical analysis, it is instructive to compare the
origin of Higgs sector CP violation in the $U(1)'$ models to that
within the MSSM. Let us first consider the case of the MSSM, in which the 
Higgs sector consists of the two electroweak Higgs doublets $H_{u,d}$.  
It is useful to start with the most general renormalizable Higgs 
potential for a two Higgs doublet model (2HDM), which 
must be built out of the gauge invariant combinations 
$|H_u|^2$, $|H_d|^2$, and $H_u\cdot H_d$ as follows:
\begin{eqnarray}
\label{2HDM}
V^{2HDM}_{ren}&=&m_{u}^{2} |H_u|^2 + m_{d}^{2} |H_d|^2 + \left( m^2_3 H_u 
\cdot H_d + \mbox{h.c.}\right)\nonumber\\
&+& \lambda_1 |H_u|^4+ \lambda_2 |H_d|^4+ \lambda_3 |H_u|^2|H_d|^2 + 
\lambda_4|H_u\cdot  H_d|^2
\nonumber\\
&+&\left [\lambda_5(H_u\cdot H_d)^2- 
\left(\lambda_6|H_d|^2+\lambda_7|H_u|^2 \right )H_u\cdot H_d + h.c.\right 
],
\end{eqnarray}
in which $m_3^2$, $\lambda_{5,6,7}$ can be complex. In a
general 2HDM, the Higgs sector exhibits CP violation if any two of these
couplings have nontrivial relative phases. 
Spontaneous CP violation can also occur for certain ranges of the 
parameters \cite{Pomarol:1992bm}. However, 
at tree level the MSSM is a special 2HDM, with $m^2_3=B\mu\equiv b$, 
$m^2_{u,d}=m^2_{H_{u,d}}$, and 
\begin{eqnarray}
\lambda_{1}=\lambda _{2}=G^{2}/4;\; 
\lambda_{3}=(g^{2}_{2}-g^{2}_{Y})/4;\;\;
\lambda _{4}=-g_{2}^{2}/2;\;\; \lambda _{5}=\lambda _{6}=\lambda
_{7}=0. 
\end{eqnarray}
As previously discussed, there is only one complex coupling $B\mu$  in the 
MSSM Higgs potential at tree level, and hence its phase can always be 
eliminated by a suitable PQ rotation of the Higgs fields.
Although the Higgs sector is CP-conserving at tree level, CP violation   
occurs at the loop level if $\theta _{f}$ and/or $\theta_{a}$ are 
nonzero, with the dominant contribution involving $\theta_{t}$. 
If $\theta_{t}\neq 0$, a relative phase $\theta$ between the VEVs of
$H_u$ and $H_{d}$ is generated \cite{pilaftsis}.

Essentially, while the $U(1)_{PQ}$ symmetry of the MSSM forbids nonzero 
values of $\lambda_{5,6,7}$ at tree level, these couplings are generated 
by radiative corrections because $U(1)_{PQ}$ is softly broken by the 
$B\mu$ term. For example, the effective $\lambda_5$ coupling which is 
generated at one-loop is approximately 
\begin{equation}
\lambda_5\sim \frac{h^{2}_{t}}{16\pi^{2}m_{SUSY}^{4}}(\mu A_{t})^{2};
\end{equation}
see \cite{pilaftsis} for the explicit expressions.\footnote{Note that 
spontaneous CP violation (SCPV) requires $m^{2}_{3}<\lambda_5 v_uv_d$. 
As $\lambda_5$ is loop suppressed in the MSSM, SCPV would require a 
very small $m^2_3$, leading to an unacceptably light pseudoscalar Higgs 
mass \cite{Pomarol:1992bm}.}

Within the $U(1)'$ models, the tree level Higgs potential does not allow 
for explicit or spontaneous CP violation.  However, it is possible to make 
a stronger statement:  unlike the MSSM, the Higgs potential in this class 
of $U(1)'$ models does not allow for CP violation at the renormalizable 
level at any order in perturbation theory. To see 
this more clearly, consider the most general renormalizable Higgs 
potential for $H_u$, $H_d$, and $S$. The potential can be expressed as a 
function of the gauge-invariant quantities $|H_u|^2$, $|H_d|^2$, 
$|H_d\cdot H_u|^2$, and $S H_u\cdot H_d$:
%***********
\begin{eqnarray}
\label{renormpot}
V_{ren}&=&m_{u}^{2} |H_u|^2 + m_{d}^{2} |H_d|^2 + 
m_S^{2}|S|^2+ \left( m_{12} S H_u \cdot H_d + 
\mbox{h.c.}\right)\nonumber\\
&+& \lambda_u |H_u|^4+ \lambda_d |H_d|^4+ \lambda_s |S|^4 \nonumber\\
&+& \lambda_{u d} |H_u|^2|H_d|^2 + \lambda_{u s}|H_u|^2|S|^2+
\lambda_{d s}|H_d|^2|S|^2 + \widetilde{\lambda}_{u d}
|H_u\cdot  H_d|^2,
\end{eqnarray}
At tree level, the dimensionful parameters $m_{u,d}^2=m^2_{H_{u,d}}$ and
$m_{12}=h_s A_s$, and the dimensionless couplings have all been
listed before except $\widetilde{\lambda}_{u d}=\frac{1}{2} g_2^{2}- h_s^{2}$.
Therefore, even in the most general renormalizable Higgs potential there 
is only one coupling which can be complex ($m_{12}$); this is because the 
gauge-invariant operator $S H_{u}\cdot H_d$ is already dimension 3.  
Hence, the global phases of the Higgs fields (more precisely of the 
combination $S H_u H_d$) can always be chosen such that the phase of 
$m_{12}$ is absorbed. Note that this statement, while true for the 
tree-level potential of (\ref{treepot}), does not depend in any way on 
perturbation theory.\footnote{Note that the structure of the potential is 
very different in the case of the NMSSM, in which the $S$ is a total gauge 
singlet. As gauge invariance then does not restrict the possible 
$S$ couplings, the Higgs sector generically violates CP at tree 
level \cite{nmssmcp}.}

As the Higgs potential conserves CP to all orders at the renormalizable
level, CP violation can enter the theory only through 
loop-induced nonrenormalizable operators. The form of 
(\ref{colemanweinberg}) demonstrates that the one-loop 
contributions to the Higgs potential include an infinite series of terms 
involving powers of the Higgs fields. While these terms include 
contributions to the potential at the renormalizable level, they also 
include a tower of nonrenormalizable terms, such as  
\begin{equation}
\label{nr1}
V_{nr}=\ldots +\left( \frac{\lambda }{m^{2}_{SUSY}}(S H_{u}\cdot
H_{d})^{2}+h.c.\right) +\ldots, 
\end{equation}
in which $m_{SUSY}$ denotes a typical sfermion mass. By $U(1)_{R}$ 
invariance, the coupling $\lambda$ of the $(S H_u\cdot H_d)^{2}$ term is 
proportional to $\lambda \sim A^{2}_t/(16\pi ^{2}m_{SUSY}^{2})$. Such  
a term is generated by the one-loop diagram formed from the Lagrangian 
interactions $h_s h^*_t S H^0_d\widetilde{u}^*_L\widetilde{u}^{c*}_L+{\rm 
h.c.}$ (from F terms) and the soft SUSY breaking interaction $h_t A_t 
\widetilde{u}_L \widetilde{u}^c_L+{\rm h.c.}$. For $\langle S\rangle \gg 
\langle H_{u,d}\rangle$, (\ref{nr1}) effectively leads to the coupling 
\footnote{Note that SCPV is also not viable in this 
potential, for the same reason as in the MSSM.}
\begin{equation}
\frac{\lambda ^{eff}_{5}}{m^{2}_{SUSY}}(H_{u}\cdot 
H_{d})^2,
\end{equation}
with 
\begin{equation}
\lambda ^{eff}_{5}\sim \frac{(A_{t}\langle S\rangle)^{2}}{16\pi^{2} 
m_{SUSY}^{2}}.
\end{equation}
In general, one can expand the one-loop potential in powers of the 
phase-sensitive gauge-invariant operator $S H_{u}\cdot H_d$:
\begin{eqnarray}
\label{deltavp}
\delta V&=&\ldots-\beta_{h_t} h_s\ {\cal{F}}\left(Q^2,
m_{\widetilde{t}_{1}}^{2}, m_{\widetilde{t}_{2}}^{2}\right)\ A_t\ S
H_{u}\cdot H_d \nonumber\\ &+& \beta_{h_t} h_t^2 h_s^2\
{\cal{G}}\left(m_{\widetilde{t}_{1}}^{2},
m_{\widetilde{t}_{2}}^{2}\right)\
\frac{A_t^2}{\left(m_{\widetilde{t}_{1}}^{2}-
m_{\widetilde{t}_{2}}^{2}\right)^{2}}\
\left(S H_{u}\cdot H_d\right)^2 + \mbox{h.c.}+\ldots,
\end{eqnarray}
in which we have presented only the phase-sensitive corrections up to
quadratic order (this expansion can of course be continued to higher
orders with no difficulty at all). The first term renormalizes the $h_s
A_s S H_{u}\cdot H_d$ operator in the tree level potential, while the 
second term is a new higher-dimensional operator. Both terms
violate CP through the phase of $A_t \langle S \rangle$ (recall this phase
is irremovable if $A_t$ and $A_s$ have a nontrivial relative phase
$\theta_{t\,s}$). The effective theory at scales below $\langle S\rangle$
is equivalent to the MSSM (with $\mu$ and $B\mu$ parameters related to the
other soft parameters of the model). One concludes from (\ref{deltavp})
that the size of the CP violation in the Higgs sector depends on the 
extent to which the $U(1)'$ breaking scale is split from the electroweak 
scale. Indeed, below the scale $\langle S\rangle$, the coefficients of the
CP-violating effective operators in (\ref{2HDM}) grow with $|A_t| v_s$ (or
equivalently $|A_t| |\mu_{eff}|$), in agreement with the CP-violating
$M_{(u,d,s)A}^{2}$ entries of the Higgs mass-squared matrix.

\section{Phenomenological Implications}
In this section we discuss the existing constraints on $U(1)'$ models
as well as their phenomenological implications with explicit CP violation.

\subsection{Constraints from $Z-Z'$ Mixing}
In the previous section, we computed the radiatively corrected Higgs boson 
mass-squared matrix (\ref{higgsmassmat}). If the eigenvalues of the Higgs 
mass-squared are all positive definite, the parameter space under concern 
corresponds to a minimum of the potential.  The parameter space is of 
course also constrained by the fact that direct collider searches have 
yielded lower bounds on the sparticle and Higgs masses. Within $U(1)'$ 
models, further constraints arise from the nonobservation to date of a 
$Z'$, both from direct searches \cite{Abe:1997fd} and indirect precision
tests from $Z$ pole, LEP II and neutral weak current data 
\cite{cveticlangacker,indirect}. The strongest constraints arise 
from the mixing mass term between the $Z$ and the $Z'$ induced by 
electroweak breaking:
\begin{eqnarray}
\label{mzzp}
M_{Z-Z'}=\left(\begin{array}{cc} M_{Z}^{2} & \Delta^{2}\\\\
\Delta^{2} & M_{Z'}^{2}\end{array}\right),
\end{eqnarray}
in which 
\begin{eqnarray}
M_{Z}^{2}&=&G^{2} v^2/4\\ 
M_{Z'}^{2}&=&g_{Y'}^{2}\left(Q_u^{2}v_u^{2} + Q_d^{2} 
v_d^{2} + Q_s^{2} v_s^{2}\right)\\ 
\Delta^{2}&=&\frac{1}{2}g_{Y'} G \left(Q_u v_u^{2} - Q_d
v_d^{2}\right). 
\end{eqnarray}
Current data requires $\Delta^{2}\ll M_{Z'}^{2}, 
M_{Z}^{2}$, because the $Z-Z'$ mixing angle 
\begin{eqnarray}
\label{alphaz}
\alpha_{Z-Z'} = \frac{1}{2} \arctan\left(\frac{2
\Delta^{2}}{M_{Z'}^{2}-M_{Z}^{2}}\right).
\end{eqnarray}
must not exceed a ${\rm few} \times 10^{-3}$ in typical models. 

Let us review the implications of this constraint, which was studied in 
\cite{cveticlangacker,Cvetic:1997ky}.  One can see from (\ref{alphaz}) 
that unless $M_{Z'}\gg M_Z$, the $Z-Z'$ mixing angle is naturally of 
$O(1)$. Therefore, a small $\alpha_{Z-Z'}$ requires a cancellation in the 
mixing term $\Delta^2$ for a given value of $\tan\beta$.  For models in 
which $M_{Z'}\sim O(M_Z)$, this cancellation 
must be nearly exact; this can be slightly alleviated if the $Z'$ 
mass is near its natural upper limit of a few TeV.  Hence, $\tan^2\beta$ must 
be tuned around $Q_d/Q_u$ with a precision determined by the size of 
${\alpha_{Z-Z'}}$. In our analysis, we will eliminate $\tan\beta$ from 
(\ref{alphaz}) for a given value of $\alpha_{Z-Z'}$:
\begin{eqnarray}
\label{tanbeta}
\tan^{2}\beta=\frac{\eta\,{Q_d} - {\alpha_{Z-Z'}}\,
      \left( -1 + \eta^2\,\left( {{Q_d}}^2 + 
           {{Q_s}}^2\,\frac{v_s^2}{v^2} \right)  \right) }{\eta\,
      {Q_u} + {\alpha_{Z-Z'}}\,
      \left( -1 + \eta^2\,\left( {{Q_u}}^2 + 
           {{Q_s}}^2\,\frac{v_s^2}{v^2} \right)  \right) },
\end{eqnarray}
in which $\eta= 2 g_{Y'}/G$, and we used
$\tan(2\, {\alpha_{Z-Z'}}) \approx 2\, {\alpha_{Z-Z'}}$.
Having fixed $\tan\beta$ in this way, a multitude of parameters remain 
which can be varied continuusly as long as all collider constraints are 
satisfied. In \cite{Cvetic:1997ky}, two phenomenologically viable 
scenarios were identified:
\begin{itemize} 
\item{\it Light $Z'$ Scenario}.  
Clearly, the $U(1)'$ symmetry can be broken along with the SM gauge 
symmetries at the electroweak scale.\footnote{As shown in 
\cite{Cvetic:1997ky}, at tree level a light $Z'$ boson with a 
vanishing $Z-Z'$ mixing (for $Q_u=Q_d$) naturally arises when 
$|A_s|$ is the dominant soft mass in the Higgs potential. Such 
trilinear coupling induced minima can also accommodate a heavy $Z'$ 
boson. This can happen in models in which there are several 
additional singlets in a secluded sector coupled to the Higgs
fields $H_{u,d}$ and $S$ via the gauge or gravitational interactions
\cite{moreU1}. Furthermore, these large trilinear coupling 
scenarios (with light $Z'$ bosons) also have interesting implications 
for baryogenesis, due to the first order phase transition at tree 
level.  If the phase transition remains first order 
after radiative corrections are included, then $\bar{\theta}$ may be 
sufficient to generate the baryon asymmetry. The electroweak phase
transition in $Z'$ models with a secluded sector is strongly
first order (with a heavy enough $Z'$ without any fine-tuning),
and electroweak baryogenesis in such models can be viable in a 
greater region of parameter space than in the minimal model 
\cite{workinprog}.} In this 
case $v_s\sim v$, 
$\tan\beta\sim \sqrt{|Q_d|/|Q_u|}$,
and $M_{Z'}$ is of order $M_{Z}$ (the precise factor depends on the size 
of $g_{Y'} |Q_s|$). However, the collider constraints on such a light $Z'$ 
are severe within typical models, and hence it can be accommodated in the 
spectrum only if it is sufficiently leptophobic. Note that 
within this framework, leptophobic $U(1)'$ couplings lead to a generic 
difficulty related to lepton mass generation: as $Q_{H_d}\neq 0$, if the 
electron mass is induced via the Yukawa coupling $h_{e} \widehat{L}_{1} 
\widehat{H}_d \widehat{E}^c_{1}$, the leptons necessarily have 
nonvanishing $U(1)'$ charges.  The electron mass (and perhaps all light 
fermion 
masses) then must be generated via nonrenormalizable 
interactions which guarantee the neutrality of $\widehat{L}_{1}$ and 
$\widehat{E}^c_{1}$ under the $U(1)'$. In practice, this 
would need to be investigated within specific models.
\footnote{However, the kinetic mixing between the hypercharge
and $Z'$ gauge bosons can be used to decouple leptons from $Z'$
though all leptons, with nonzero $U(1)'$ charges, acquire 
their masses from their Yukawa couplings \cite{kinmix}.}

\item{\it Heavy $Z'$ Scenario.} 
In this scenario, the $U(1)'$ breaking is radiative (driven by the 
running of $m_S^2$ to negative values in the infrared) and occurs at a 
hierarchically larger scale than the electroweak scale. However, 
gauge invariance does not allow for the $U(1)'$ and electroweak breakings 
to decouple completely (as $Q_{H_{u,d}}\neq 0$). The electroweak scale is 
then achieved by a cancellation among the soft masses, which are typically 
of $O(M_{Z'})$, with a fine-tuning $O(M_{Z'}/M_Z)$. As 
discussed in \cite{Cvetic:1997ky}, excessive fine tuning is avoided if 
$M_{Z'}$ in units of the heavy scale is roughly bounded by the ratio of 
the charges, $\mbox{Min}[|Q_s/Q_d|,|Q_s/Q_d|]$. There are several 
advantages of the heavy $Z'$ scenario.  First, the $Z-Z'$ mixing can be 
kept small enough with less fine-tuning of the $\Delta^2$ in 
(\ref{mzzp}); in particular, $Q_u=Q_d$ is no longer a requirement. In 
addition, the collider constraints are less severe for $Z'$ bosons with 
TeV-scale masses in typical models; for example, leptophobic couplings are 
not generically a phenomenological necessity.

\end{itemize}

\subsection{Constraints from Dipole Moments}
Let us now turn to dipole moment constraints.  Recall in SUSY theories 
dipole moments of the fundamental fermions are generated by 
gaugino/Higgsino exchanges accompanied by sfermions of the appropriate 
flavor.  The dipole moment under concern may (e.g. the electric
and chromoelectric dipole moments of the quarks) or may not (e.g. the 
anomalous magnetic moment of muon) require explicit sources of CP 
violation. 

In the MSSM, dipole moments can provide important constraints on the
parameter space. For example, the anomalous magnetic moment of the muon is
in principle an important observable either for discovering SUSY
indirectly or constraining SUSY parameter space; however, at present the
theoretical uncertainties present in certain nonperturbative SM
contributions lead to difficulties in carrying out this procedure using
recent data (see e.g. \cite{marciano} for a review of the basic physics
and \cite{muongm2exp} for the most recent experimental results). At
present, the most stringent constraints arise from electric dipole moments
(EDMs). As is well known, the experimental upper bounds on the EDMs of the
electron, neutron, and certain atoms impose particularly severe
constraints on the parameter space of general SUSY models.  In contrast to
the SM, in which EDMs are generated only at three-loop order (as the only
source of CP violation is in the CKM matrix), the sources of explicit CP
violation in SUSY theories include phases in flavor-conserving couplings
which, if present, lead to nonvanishing one-loop contributions to the EDMs
which can exceed the experimental bounds.  As these phases generically
filter into the Higgs sector, it is important to include the parameter
space constraints provided by the EDM bounds.

\begin{figure}
\vspace*{-0.25in}
\begin{center}
%\begin{minipage}{8in}
\epsfig{file=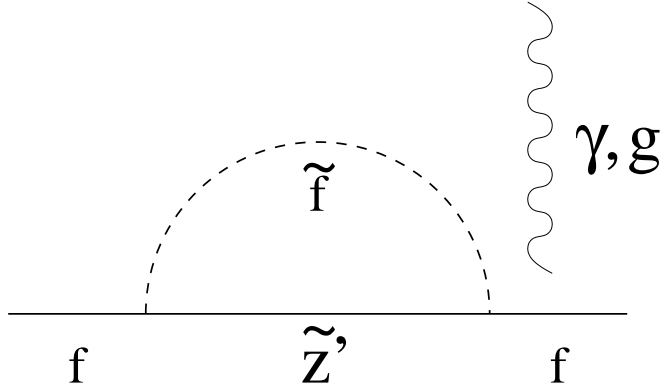,height=2in}
%\end{minipage}  
\end{center}
%\vskip .1in
\caption{\label{fig0}
{\footnotesize The $\widetilde{Z'}$--sfermion diagram which 
contributes to the (chromo-)electric and (chromo-)magnetic dipole
moments of the fermion $f$. The photon ($\gamma$) or
gluon ($g$) are to be attached in all possible ways.}} 
\end{figure}

Let us consider the dipole moments which arise within this class of 
$U(1)'$ models. After replacing the $\mu$ parameter 
of the MSSM by $\mu_{eff}$ in (\ref{mupar}), all one-loop dipole 
moments are found to be identical to their MSSM counterparts except for 
an additional contribution generated by the 
$\widetilde{Z'}$--$\widetilde{f}$ diagram in Fig.\ref{fig0}. Here 
$\widetilde{Z'}$ is the $U(1)'$ gaugino with mass $M_{1'}$.
This diagram generates the operator $D_{f}\overline{f_L} 
\sigma^{\mu\nu}F_{\mu \nu} f_R$, in which  
\begin{eqnarray}
\label{df}
D_{f}(\widetilde{Z'})\sim \frac{g_{Y'}^{2} Q_f^2} {16 \pi^2}\ 
\frac{m_f\ |M_{1'}|}{M_{\widetilde{f}}^{4}}\ \left[ |A_f|\ 
e^{i(\theta_{1' s} - \theta_{f s})}
- R_f |\mu_{eff}|\ e^{i(\theta_{1' s} + \overline{\theta})}\right],
\end{eqnarray}
in units of the electromagnetic or strong coupling. In 
the above $R_f=(\tan\beta)^{-2 
I^3_f}$, $M_{\widetilde{f}}$ characterizes the typical sfermion 
mass, and recall that the reparameterization
invariant phases are defined in (\ref{u1pphases}). Clearly, the 
(chromo-)electric and (chromo-)magnetic dipole moments are generated, 
respectively, by $\mbox{Im}[D_{f}]$ and $\mbox{Re}[D_{f}]$.
The expression above is approximate estimate (valid in the limit that 
$M_{\widetilde{f}}\gg M_{1'}$) of the exact amplitude; a more precise
treatment would take into account the mixing of all six neutral fermions. 
The amplitude (\ref{df}) is similar to the bino exchange contribution 
in the MSSM. 

Within the aforementioned light $Z'$ scenario, for phenomenologically
viable models $D_{e}(\widetilde{Z'})$ vanishes because the lepton 
couplings to the $Z'$ are necessarily leptophobic. Therefore, for 
instance, the electron EDM is completely decoupled from the presence 
of an electroweak scale $U(1)'$ symmetry. This conclusion extends to 
other leptons 
for family universal $Z'$ models.  This may also be relevant for the 
hadronic dipole moments depending on whether or not the $Z'$ boson is 
hadrophobic (assuming it is detected in present and/or forthcoming 
colliders). As $|\mu_{eff}|\ll M_{\widetilde{f}}$ within the light $Z'$ 
scenario, the dipole moments of both the up-type and down-type fermions 
are largely controlled by the corresponding $A_f$ parameters. 
In contrast, the $U(1)'$ charges are not necessarily suppressed for any 
fermion flavor in the heavy $Z'$ scenario and thus the 
$D_{f}(\widetilde{Z'})$ contribution to dipole moments can compete with 
the MSSM amplitudes. In this scenario, $|\mu_{eff}|\sim M_{Z'}\gg M_Z$, 
and hence both terms in $D_{f}(\widetilde{Z'})$ are important. The dipole 
moments become sensitive to $\theta_{1' s} + \overline{\theta}$ if the 
$A_f$ parameters are sufficiently small compared to $|\mu_{eff}|$. 

As the dipole moments generically scale as $m_f/M_{\widetilde{f}}^{2}$, 
(which is clear from the form of (\ref{df})), when 
$M_{\widetilde{f}} \sim O(M_Z)$ the EDMs typically exceed the existing 
bounds by 2 to 3 orders of magnitude if the phases are $O(1)$. As 
discussed briefly in the Introduction, one possibility is satisfying the 
experimental bounds while retaining $O(1)$ phases is to raise 
$M_{\widetilde{f}}$ to multi-TeV values, which in 
effect requires the sfermions of first and second generations to be 
ultraheavy \cite{oldedm,newedm}. Another way of suppressing the EDMs is to 
invoke accidental cancellations between different contributions, {\it 
i.e.} to find regions of parameter space in which the SUSY amplitudes 
interfere destructively. In the MSSM with low values of $\tan\beta$ , this 
has been shown to occur with almost no costraint on any of the invariant 
phases except 
$|\theta_{\widetilde{\chi}^{\pm}}|=|\phi_{\mu}+\phi_{M_2}|\simlt \pi/10$ 
\cite{newedm,ibrahimnath,Brhlik:1998zn,Pokorski:1999hz,Barger:2001nu}.
This strong constraint follows from the fact that $g_{Y}\ll g_2$, and thus 
the $SU(2)$ gauginos dominate the EDMs. Within the $U(1)'$ framework, 
%which  allows for a dynamical solution to the $\mu$ problem of the MSSM, 
the EDM constraints can have varying implications depending on the size of 
(\ref{df}).
\begin{itemize}

\item If $g_{Y'}\sim O(g_Y)$ or (more generally) $g_{Y'}\ll g_2$, 
the EDM constraints on the parameter space are similar to that of the 
MSSM except for a slight folding of the cancellation domain due to the 
inclusion of (\ref{df}). Once again, the most strongly constrained phase 
is $\theta_{\widetilde{\chi}^{\pm}}$, with 
$|\theta_{\widetilde{\chi}^{\pm}}|\simlt \pi/10$ in the low $\tan\beta$ 
regime. As $\theta_{\widetilde{\chi}^{\pm}}=\theta_{2s}+\overline{\theta}$ 
and $\overline{\theta}$ is a loop-suppressed angle (\ref{theta}), the EDMs 
provide a constraint on $\theta_{2s}$: $|\theta_{2s}| \simlt \pi/10$. 
Consequently, the dynamical solution to the $\mu$ problem present in 
this class of $U(1)'$ models also solves the SUSY CP hierarchy problem in 
specific models of the soft parameters in which (at least) the $SU(2)$ 
gaugino mass has the same phase as the $A_s$ parameter (then 
$\theta_{2s}=0$ by definition).\footnote{In this 
paper, we have not addressed the origin of the phases of the soft 
parameters in (\ref{soft}), and hence we cannot make any claims about how 
one solves the SUSY CP hierarchy problem within this framework.  However, 
it is worthwhile to note that models of the soft parameters in which the 
gaugino masses and $A$ terms have the same phases are quite common within 
various classes of four-dimensional string models (at least at tree level) 
under plausible assumptions \cite{ibanez}.}

\item If $g_{Y'} \simgt g_2$,  the dipole moment 
amplitude $D_{f}(\widetilde{Z'})$ becomes comparable to or larger than the 
$SU(2)$ gaugino contribution, and the cancellation domain found in the 
MSSM will be significantly folded. In this case, the EDMs will constrain
a combination of the phases in (\ref{df}) and 
$\theta_{\widetilde{\chi}^{\pm}}$. Such a scenario, however, can have 
tension with the standard picture of gauge coupling unification at a high 
fundamental scale (although in principle it could be considered as a 
possibility in generic low scale realizations).
\end{itemize}

Until this point, we have only discussed one-loop EDMs.  It was pointed
out a while ago \cite{twoloop} that in certain regions of MSSM parameter
space certain two-loop contribution contributions which exclusively depend
on the third generation sfermions can be nonnegligible. These
contributions, which are particularly relevant if the one-loop EDMs are
suppressed solely by ultraheavy first and second generation sfermion
masses, involve the same phases which predominantly filter into the Higgs
potential at one-loop ({\it i.e.} the phases present in the stop
mass-squared matrix). However, these two-loop EDMs become sizeable only at
large $\tan\beta$. In this paper, we have restricted our attention to
small $\tan\beta$ values, which is a well-motivated parameter regime
(e.g., $\tan\beta=1$ is allowed within this framework, in contrast to the
MSSM). Hence, these contributions will not provide significant parameter
space constraints in our numerical analysis.

\subsection{Numerical Estimates for Higgs Sector CP Violation} 
In this section, we present sample numerical calculations of the Higgs
boson masses and mixings derived in Section 3.2, taking into account the
phenomenological constraints on the parameter space discussed in Sections 
4.1 and 4.2.

In the absence of CP violation, the scalar-pseudoscalar mixing
terms of the Higgs mass-squared matrix (\ref{higgsmassmat}) vanish
($\sin\bar{\theta}=0$), and hence (\ref{higgsmassmat}) takes on a block
diagonal form.  The structure of (\ref{higgsmassmat}) demonstrates that in
this limit there is one CP even scalar with mass $\propto v_s$ and a CP
odd scalar with mass proportional to $\sqrt{|A_s| v_s}$. In addition,
there is a light CP even scalar of mass $\sim M_Z$ and a heavier CP even
scalar with its mass controlled by a combination of $v$ and $M_A$.
However, in the presence of explicit CP violation, the Higgs bosons cease 
to have definite CP parities. The strength of CP violation in the Higgs 
sector is parameterized by the reparameterization invariant phase 
$\theta_{t\, s}$, which induces a nonvanishing $\bar{\theta}$  through the 
relation (\ref{theta}). The induced phase $\overline{\theta}$ is a 
loop-induced and scale-dependent quantity which is particularly enhanced 
in parameter regions with a low $M_A$.

As discussed in Section 4.2, while the one-loop EDM constraints strongly 
constrain the phase $\theta_{2s}$, this phase is not the dominant source 
of CP violation in the Higgs sector for small values of $\tan\beta$ and 
hence this constraint does not restrict the parameter space for our 
analysis.  The dominant corrections to the Higgs potential arise from 
top and stop loops, and the dipole moments of the fermions in first two 
generations feel such effects only at two loop level. In fact, in low 
$\tan\beta$ limit (which is the domain in which our analysis of the Higgs 
potential is valid), such effects are completely negligible 
\cite{twoloop}. Therefore, the EDM constraints do not have a direct 
impact on our analysis of CP violation in the Higgs sector (we 
simply assume that the dipole moment constraints have been saturated 
either via cancellations or by choosing the first and second generation 
sfermion masses heavy enough; we could also simply assume that all 
phases except $\theta_{t\,s}$ are small). 

We now turn to the analysis of the parameter space, including the
nontrivial constraints arising from $Z-Z'$ mixing.  The fundamental
parameters relevant for the Higgs sector include $\{v_s$, $A_s,$ $A_t,$ 
$M_{\widetilde{Q}},$ $M_{\widetilde{U}}$,$h_s$, $Q_u,$ 
$Q_d,$ $g_{Y'}$, $\theta_{t\, s}\}$. We fix a
subset of these parameters as follows: (i) $\alpha_{Z-Z'}= 10^{-3}$, which
is well below the present bounds; (ii) $g_{Y'}^{2}=(5/3) G^2
\sin^{2}\theta_{W}$, as inspired from one-step GUT breaking; (iii)  
$h_s=1/\sqrt{2}$, as motivated by the RGE analysis of
\cite{Cvetic:1997ky}; (iv) $Q_u=Q_d=-1$, such that $\tan\beta$ remains
close to unity (as can be seen from (\ref{tanbeta})); and finally (v)  
$M_{\widetilde{Q}}=M_{\widetilde{U}}$. The remaining parameters can be
fixed on a case by case basis depending on the range of values assumed for
$M_{Z'}$. A few notational comments are also in order. Although 
(\ref{higgsmassmat}) suggests that $M_A$ can be chosen to be a fundamental 
parameter and this is what is traditionally done in the MSSM, we prefer to 
work instead with $A_s$ for consistency with previous discussions in this 
paper as well as the tree level analysis of \cite{Cvetic:1997ky}. In 
addition, in our numerical results we fix the renormalization scale to be 
$Q=(2 m_t + M_{Z'})/2$.  This differs once again from the MSSM, where the
renormalization scale is chosen to be $Q=m_t$ in order to minimize the
next-to-leading order corrections. Such higher order corrections are
beyond the scope of this paper; our choice for $Q$ can be regarded as some
nominal value in between the electroweak and $U(1)'$ breaking scales.

\begin{figure}
\vspace*{-0.75in}
\hspace*{-0.2in}
\begin{minipage}{10in}
\epsfig{file=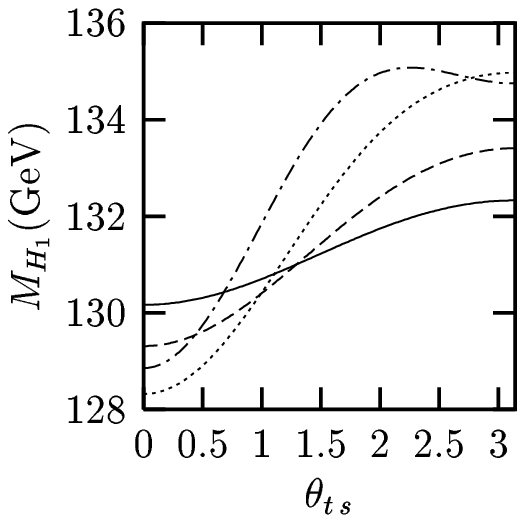,height=3.25in}
\hspace*{0.5in}
\epsfig{file=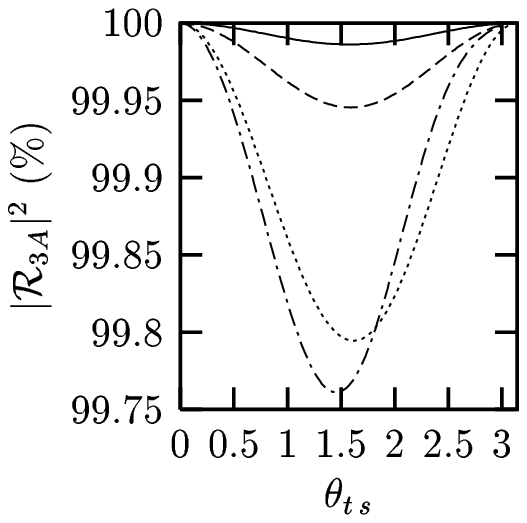,height=3.25in}
\end{minipage}
\vskip .2in
\caption{\label{fig1}
\footnotesize The $\theta_{t\, s}$ dependence of the lightest Higgs
mass and the CP-odd composition of $H_3$ in
the light $Z'$ scenario. The solid, 
dashed, dotted, and dot-dashed curves correspond to, respectively,
$|A_t|/v_s=1/2, 1, 2$ and $4$ with $v_s=v/\sqrt{2}$.}
\end{figure}

We begin with an analysis of the light $Z'$ scenario. For purposes of 
definiteness, we set $M_{\widetilde{Q}}= 2 v_s$, $v_s = v/\sqrt{2}\simeq 
m_t$, and $|A_s|=v_s$, in which case $M_{Z'}\simeq 2 M_Z$ and 
$|\mu_{eff}|\simeq M_Z$.  The SUSY phase $\theta_{t\, s}$ influences both 
the Higgs masses and their mixings, as shown in Figure \ref{fig1}.  In the 
left panel, the variation of the lightest Higgs mass with $\theta_{t\, s}$ 
is displayed for several values of $|A_t|/v_s$. For $|A_t|/v_s= 
1/2, 1$ and $2$ $M_{H_1}$ grows gradually with $\theta_{t\, s}$, peaking 
at $\theta_{t\, s} =\pi$.  This behaviour is easy to understand: as 
the magnitude of the stop LR mixing depends strongly on $\theta_{t\, s}$, 
the variation of $M_{H_1}$ with respect to $\theta_{t\, s}$ simply 
displays the well-known fact that the lightest Higgs mass depends strongly 
on the value of the stop mixing. Indeed, 
\begin{equation}
\label{stopLRratio}
\frac{|M_{LR}^{2}|_{\theta_{t\, s} = \pi}}{|M_{LR}^{2}|_{\theta_{t\, s} = 
0}}= \frac{|A_t| + |\mu_{eff}|\cot\beta}{|A_t| - |\mu_{eff}|\cot\beta},
\end{equation} 
which becomes large when $|A_t|$ and $|\mu_{eff}|$ are of comparable size. 
The ratio (\ref{stopLRratio}) gets saturated with further increase 
of $|A_t|$; however, in this case $|A_t|  |\mu_{eff}|$ also becomes large, 
which affects both the $M_{P}^{2}$ and $M_{i A}^{2}$ entries of the Higgs 
mass-squared matrix. While the former shifts the peak value of $M_{H_1}$ 
towards the point of maximal CP violation (see the dot-dashed
curve in the figure), the latter enhances the scalar-pseudoscalar
mixings. The generic strength of the scalar-pseudoscalar mixings can be
determined e.g. by working out the CP-odd composition
of $H_3$ (the would-be pseudoscalar Higgs). The result is
shown in the right panel of Figure \ref{fig1}. Clearly,
the $M_{i A}^{2} \sin\theta_{t\, s}$ elements of the
Higgs mass-squared matrix are not large enough to enhance
such mixings ($|{\cal{R}}_{3 A}|^{2}$ falls at 
most to $99.75\%$ for $|A_t| = 4 v_s$). 

The functional dependence of the heavier Higgs boson masses on 
$\theta_{t\, s}$ is opposite that of $M_{H_1}$ in that the masses tend to 
decrease as $\theta_{t\, s}$ ranges from 0 to $\pi$; e.g. when  
$|A_t| = 4 v_s$, $(M_{H_4}, M_{H_3}, M_{H_2})$ fall from ($245, 224, 191$) 
to ($234, 206, 182$) ${\rm GeV}$. In accord with the analytical expression 
(\ref{theta}), $\overline{\theta}$ grows with $|A_t|$ until it arrives at 
the peak value of $\sim 30\%$ for $|A_t| = 4 v_s$ for maximal CP 
violation. For low $M_{Z'}$ minima, the scalar-pseudoscalar mixings (which 
govern the novel CP violating effects in the Higgs couplings to fermions, 
gauge bosons and other Higgs bosons) are typically small due to the low 
value of 
$|\mu_{eff}|$. 

\begin{figure}
\vspace*{-0.75in}
\hspace*{-0.2in}
\begin{minipage}{10in}
\epsfig{file=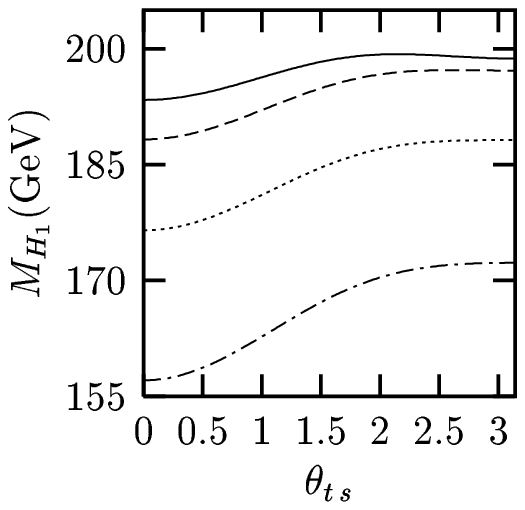,height=3.25in}
\hspace*{0.5in}
\epsfig{file=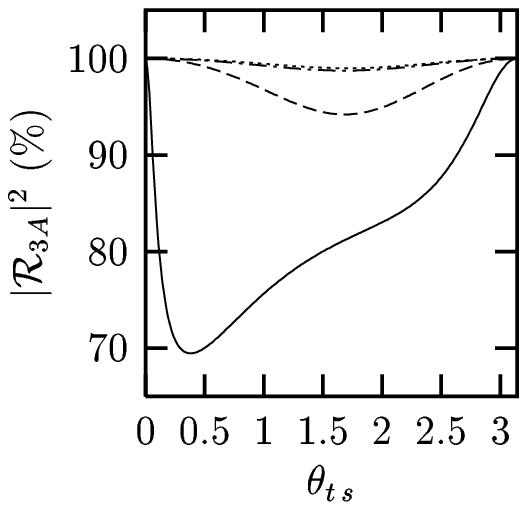,height=3.25in}
\end{minipage}
\vskip .2in
\caption{\label{fig2}
\footnotesize The $\theta_{t\, s}$ dependence of the lightest Higgs
mass (left panel) and the CP-odd composition of $H_3$ in the heavy
$Z'$ scenario. Here solid, dashed, dotted, and dot-dashed curves 
correspond, respectively, to $|A_s|/v_s=1/5, 1/2, 3/4$ and $1$ with $v_s=1\ 
{\rm TeV}$.}
\end{figure}

We now discuss the heavy $Z'$ scenario, setting $v_s = 1\ {\rm TeV}$, 
$M_{\widetilde{Q}}=750\ {\rm GeV}$, and $|A_t|=2 M_{\widetilde{Q}}$. 
Figure~\ref{fig2} depicts the variations of the lightest Higgs mass (left 
panel) and the CP-odd composition of the would-be Higgs scalar
as a function of $\theta_{t\, s}$ and $|A_s|$. In both figures,
the solid, dashed, dotted, and dot-dashed curves
correspond, respectively, to $|A_s|/v_s=1/5, 1/2, 3/4$ and $1$.
In contrast to the light $Z'$ scenario as shown in Figure~\ref{fig1},  
here we illustrate the dependence on $|A_s|$ (or equivalently $M_A$), as 
this parameter remains largely free in the heavy $Z'$ limit 
\cite{Cvetic:1997ky}. As the left panel of the
figure shows, the mass of the lightest Higgs is typically 
larger than that in the light $Z'$ scenario.  The lightest Higgs mass is 
also a steep function of $|A_s|$, which becomes increasingly smaller as  
$M_A$ increases due to decoupling. Note also that the dependence of  
$M_{H_1}$ on $\theta_{t\, s}$ in this scenario is similar
to the case within the light $Z'$ scenario; once again, this is because 
the radiative corrections to the lightest Higgs mass strongly depend on 
the value of the stop mixing parameter.

However, in contrast to the light $Z'$ scenario, the scalar-pseudoscalar 
mixings in the heavy $M_{Z'}$ limit are sizeable, as shown in the 
right panel of Figure~\ref{fig2}. This feature is expected because the 
strength of the Higgs sector CP violation is governed by the size of the 
singlet VEV, {\it i.e.} the effective $\mu$ parameter, and in this 
scenario  $|\mu_{eff}| \sim M_{Z'}$.  In general, the CP-violating mixings 
grow larger as $A_s$ decreases, because in this case 
$M^2_{iA}\sin\theta_{t\,s}$ can be 
comparable to $M_A$, which facilitates scalar-pseudoscalar transitions.  
For $|A_s|=v_s/5$, the CP-odd composition of the would-be pseudoscalar 
Higgs falls down to $70 \%$ around $\theta_{t\, s}\sim \pi/6$. However, as 
$|A_s|$ increases (while keeping $|\mu_{eff}|$ and $|A_t|$ fixed), the 
diagonal elements of (\ref{higgsmassmat}) also increase, with the result 
that the CP-violating effects become weaker. The large
variations in $|{\cal{R}}_{3 A}|^{2}$ depicted here 
are due to the mixings between $H_3$ and $H_2$. Indeed,
for $|A_s|/v_s=1/5, 1/2, 3/4$ and $1$ the two masses are 
strongly degenerate, with ($M_{H_2}, M_{H_3}$)
starting at  ($476, 477$), ($722, 726$), ($876, 881$), 
($1013, 1007$) and decreasing to ($417, 418$), ($685, 688$), ($846, 851$), 
($987, 981$) ${\rm GeV}$ as $\theta_{t\, s}$ varies from 0 to $\pi$. Note 
that the scalar-pseudoscalar conversions are more efficient when the two 
masses are highly degenerate. 

For the values of $|A_s|/v_s$ exhibited above, the Higgs sector is within
the decoupling regime ($M_A > 2 M_Z$)\footnote{See
\cite{pilaftsis,refined} for a more precise definition of the decoupling
regime in the CP-violating MSSM.}, in which the lightest Higgs resembles 
the SM Higgs boson, the heaviest Higgs is singlet-dominated with a mass of 
order $M_{Z'}$, and the two intermediate mass Higgs (the CP odd scalar and 
the second heaviest CP even scalar in the absence of CP violation) are 
strongly degenerate.  The lightest Higgs
boson is essentially CP even ($|R_{1A}|^2\ll 0.1\%$ for $|A_s|/v_s\geq
1/15$) and hence is decoupled from CP-violating effects, although its mass
depends strongly on $\theta_{ts}$.  However, there
are phenomenologically interesting corners of parameter space with
sufficiently small values of $|A_s|/v_s$ in which the lightest Higgs boson
can have a significant mixing with the would-be pseudoscalar.  
As the lightest Higgs mass is a steep function of $|A_s|/v_s$, for a
value of $M_{H_1}$ consistent with LEP bounds the CP-odd composition of
$H_1$ cannot be larger than $20\%$.  It is important to keep in mind
though that the couplings of the lightest Higgs boson to gauge bosons and
fermions are modified when the lightest Higgs has a significant mixing
with the would-be pseudoscalar (the modifications grow with the CP-odd
composition of the lightest Higgs), such that the existing LEP bounds may
not be applicable (see e.g. \cite{CPX,CPXX} for discussions within the 
MSSM).

Our results demonstrate that the CP-violating effects in the
Higgs sector, or more precisely, the mixing between the would-be scalars 
and pseudoscalars in the CP conserving limit, are generically highly
suppressed in the light $Z'$ models but can be sizeable in the heavy $Z'$
scenario, even though the masses can vary strongly with $\theta_{t\,s}$
(which is of course a CP-conserving effect).  This behaviour is exactly in
accordance with the general discussion of Section 3.3, in which we
demonstrated that the CP-violating terms in the Higgs potential
necessarily originate from nonrenormalizable terms present at one-loop
(such terms are encoded within the full Coleman-Weinberg potential).  The 
strength of such terms in e.g. the doublet sector then scale 
according to the ratio of the singlet VEV $v_s\simeq |\mu_{eff}|$ to the 
scale of a typical soft mass. Hence, within the light 
$Z'$ scenario (in which the effective $\mu$ parameter is small) CP-violating 
effects are suppressed, while the large $|\mu_{eff}|$ present in the heavy 
$Z'$ scenario can allow for spectacular effects of CP violation.

We close this section with a brief discussion of the 
implications for collider searches.  In general, at least a subset of the 
Higgs masses within this class of $U(1)'$ models can be observable at 
forthcoming colliders and future colliders such as TESLA and NLC.
next generation of colliders such as TESLA and NLC. Within light $Z'$ 
models, all of the Higgs bosons remain light after including radiative 
corrections, but such models generically have very small CP-violating 
Higgs couplings.  In contrast, large $Z'$ models can have large 
CP-violating Higgs couplings. As the viable regions of space typically 
correspond to the decoupling limit in which all of the Higgs bosons 
except the lightest Higgs are heavy, detecting the CP-violating effects 
within this scenario is similar to that within the MSSM for a large 
$\mu$ parameter.
Such effects have been studied in \cite{pilaftsis,CPX,CPXX}, where it is 
known that Higgs sector CP violation can introduce sizeable modifications 
in the couplings of the Higgs bosons to fermions and vector bosons, and 
strongly affect the bounds inferred from the CP-conserving
theory. Furthermore, the CP purity of the Higgs bosons (assuming that 
the collider searches establish their existence) can be tested by 
measuring CP violation in its decays into heavy quarks or vector bosons 
\cite{top}.

\section{Summary}

In this paper, we have discussed the nature and implications of explicit
CP violating phases present in the soft breaking Lagrangian within
supersymmetric models with an additional $U(1)$ gauge symmetry and an
additional SM gauge singlet $S$.  This class of models is worthy of
further study not only because such gauge extensions are ubiquitous within
four-dimensional string models (and other plausible extensions of the
MSSM), but also they provide an elegant framework in which the
$\mu$ problem of the MSSM. The solution, which is to forbid 
the bare $\mu$ term by $U(1)'$ gauge invariance and generate an effective
$\mu$ parameter through the VEV of the singlet $S$, is similar to that
found within the NMSSM (but its generic cosmological and CP 
problems).
Our results can be summarized as follows:
\begin{itemize}

\item All reparameterization invariant phases can be expressed as linear
combinations of $\theta_{f\,s}\equiv \phi_{A_f}-\phi_{A_s}$ and
$\theta_{a\,s}\equiv \phi_{M_a}-\phi_{A_s}$ (and hence a ``natural" basis
can be obtained by using $U(1)_R$ to set $\phi_{A_s}=0$).

\item The Higgs sector is manifestly CP conserving at tree level (and 
indeed at renormalizable level to all orders in perturbation theory). 
However, the CP-violating phases present in the stop mass-squared matrix 
filter into the Higgs sector at the nonrenormizalizable level at one-loop.  
The CP-violating effects are particularly enhanced when  $U(1)'$ symmetry
is broken near the sparticle thresholds.

\item The spontaneous breakdown of the $U(1)'$ symmetry near the weak
scale stabilizes not only the modulus of $\mu$ but also its phase.  The
phase of $\mu$ itself is of course not a basis-independent quantity;
however, in the ``natural" basis defined above, this phase ($\bar{\theta}$
in this basis) arises only at the loop level and is typically 
$1$--$10 \%$, depending on the size of $M_A$ (the pseudoscalar Higgs boson 
mass in the CP conserving limit).

\item The absence of permanent EDMs for leptons and hadrons (even assuming
either cancellations and/or heavy first and second generation sfermions)
strongly bound the reparameterization invariant phase present in the
chargino mass matrix $(\phi_{\mu}+\phi_{M_2})=
(\bar{\theta}-\phi_{A_s}+\phi_{M_2})$, while the other SUSY phases remain
largely unconstrained.  In specific models in which the phase difference
between (at least the $SU(2)$)  gaugino mass parameters and $A_s$ is
vanishingly small, this ``SUSY CP hierarchy problem" is resolved because
the radiative phase $\bar{\theta}$ is sufficiently small to be easily
allowed by EDM bounds.

\item The CP-violating effects in the Higgs sector are quite distinct for
the two phenomenologically viable scenarios with acceptably small $Z-Z'$
mixing, because these effects are proprotional to the size of the
effective $\mu$ term. In scenarios with a light $Z'$, CP-violating effects
are suppressed, while heavy $Z'$ models can exhibit significant
CP-violating scalar-pseudoscalar mixings, with phenomenological
implications similar to that of the MSSM with large $\mu$ parameter.

\end{itemize}

\section{Acknowledgments}

We thank Paul Langacker for many illuminating discussions about $U(1)'$
models and for comments on the manuscript, and Keith Olive for discussions
about EDM bounds on the phase of the $\mu$ parameter.  We also thank Jose
Ramon Espinosa, Hassib Amini, Apostolos Pilaftsis, Alon Faraggi, and
Daniel Chung for helpful and informative conversations.  This work is
supported in part by the DOE grant DE-FG02-94ER40823.

\end{document}